\documentclass[10pt, aps, pra, twocolumn, superscriptaddress, floatfix, showpacs, longbibliography]{revtex4-1}

\usepackage{graphicx}
\usepackage{amsfonts}
\usepackage{amssymb}
\usepackage{amsmath}
\usepackage{txfonts}
\usepackage{lipsum}
\usepackage{color}
\usepackage{wasysym}
\usepackage{hyperref}
\usepackage{bbold}
\usepackage{comment}
\usepackage{etoolbox} 
\graphicspath{{figures/}}
\usepackage[normalem]{ulem} 
\usepackage{mathtools} 

\usepackage{mathrsfs} 

\newcommand{\av}[1]{\ensuremath{\left\langle #1 \right\rangle}}
\newcommand{\qv}{\mathbf{q}}
\newcommand{\kv}{\mathbf{k}}

\usepackage{letltxmacro}
\LetLtxMacro{\oldsqrt}{\sqrt}
\renewcommand{\sqrt}[2][\mkern8mu]{\mkern-6mu\mathop{}\oldsqrt[#1]{#2}}
\usepackage[compat=1.1.0]{tikz-feynman}

\begin{document}

\title{
Impact of partially bosonized collective fluctuations on electronic degrees of freedom
}

\author{V. Harkov}
\affiliation{I. Institute of Theoretical Physics, University of Hamburg, Jungiusstrasse 9, 20355 Hamburg, Germany}
\affiliation{European X-Ray Free-Electron Laser Facility, Holzkoppel 4, 22869 Schenefeld, Germany}

\author{M. Vandelli}
\affiliation{I. Institute of Theoretical Physics, University of Hamburg, Jungiusstrasse 9, 20355 Hamburg, Germany}
\affiliation{The Hamburg Centre for Ultrafast Imaging, Luruper Chaussee 149, 22761 Hamburg, Germany}
\affiliation{Max Planck Institute for the Structure and Dynamics of Matter,
Center for Free Electron Laser Science, 22761 Hamburg, Germany}

\author{S. Brener}
\affiliation{I. Institute of Theoretical Physics, University of Hamburg, Jungiusstrasse 9, 20355 Hamburg, Germany}
\affiliation{The Hamburg Centre for Ultrafast Imaging, Luruper Chaussee 149, 22761 Hamburg, Germany}

\author{A. I. Lichtenstein}
\affiliation{I. Institute of Theoretical Physics, University of Hamburg, Jungiusstrasse 9, 20355 Hamburg, Germany}
\affiliation{European X-Ray Free-Electron Laser Facility, Holzkoppel 4, 22869 Schenefeld, Germany}
\affiliation{The Hamburg Centre for Ultrafast Imaging, Luruper Chaussee 149, 22761 Hamburg, Germany}

\author{E. A. Stepanov}
\email{evgeny.stepanov@polytechnique.edu}
\affiliation{CPHT, CNRS, Ecole Polytechnique, Institut Polytechnique de Paris, F-91128 Palaiseau, France}

\begin{abstract}
In this work we present a comprehensive analysis of collective electronic fluctuations and their effect on single-particle properties of the Hubbard model.
Our approach is based on a standard dual fermion/boson scheme with the interaction truncated at the two-particle level.
Within this framework we compare various approximations that differ in the set of diagrams (ladder vs exact diagrammatic Monte Carlo), and/or in the form of the four-point interaction vertex (exact vs partially bosonized).
This allows to evaluate the effect of all components of the four-point vertex function on the electronic self-energy.
In particular, we observe that contributions that are not accounted for by the partially bosonized approximation for the vertex have only a minor effect on electronic degrees of freedom in a broad range of model parameters.
In addition, we find that in the regime, where the ladder dual fermion approximation provides an accurate solution of the problem, the leading contribution to the self-energy is given by the longitudional bosonic modes. 
This can be explained by the fact that contributions of transverse particle-hole and particle-particle modes partially cancel each other.
Our results justify the applicability of the recently introduced dual triply irreducible
local expansion ($\text{D-TRILEX}$) method that represents one of the simplest consistent diagrammatic extensions of the dynamical mean-field theory.
We find that the self-consistent $\text{D-TRILEX}$ approach is reasonably accurate also in challenging regimes of the Hubbard model, even where the dynamical mean-field theory does not provide the optimal local reference point (impurity problem) for the diagrammatic expansion.
\end{abstract}

\maketitle

\section{Introduction}

Long-range correlations play a crucial role in strongly interacting electronic systems. 
They are responsible for various phenomena as for instance magnetism, charge density waves and superconductivity.
A consistent treatment of nonlocal collective electronic fluctuations often appears to be a challenging task. 
It is important for an accurate description not only of these effective bosonic modes themselves, but also of their influence on the single-particle characteristics of the system.

A consistent model description of strongly-correlated materials should be able to identify leading collective instability channels governing physical processes in the system.
Apart from giving physical insight, this often drastically diminishes technical efforts required for solving the problem.
Hubbard model is a minimal model that accounts for the interplay between kinetic energy and Coulomb interaction of electrons.
For infinite number of spatial dimensions, the Hubbard model can be solved exactly by means of the dynamical mean-field theory (DMFT)~\cite{RevModPhys.68.13}, where the self-energy becomes purely local~\cite{PhysRevLett.62.324}. 
DMFT is a nonperturbative method that accurately accounts for local correlations by mapping the original lattice model onto an auxiliary local impurity problem, which can be solved numerically exactly. 
In finite dimensions DMFT turns out to be a good approximation for single-particle quantities, in particular when local correlations are strong~\cite{PhysRevB.91.235114, PhysRevX.11.011058}.
However, DMFT reaches its limits when spatial fluctuations become large~\cite{PhysRevB.102.224423}.

Further, cluster extensions of DMFT
~\cite{PhysRevB.58.R7475, PhysRevB.62.R9283, RevModPhys.77.1027, PhysRevLett.87.186401, doi:10.1063/1.2199446, RevModPhys.78.865, PhysRevB.94.125133} have been introduced to consider nonlocal correlation effects.
However, the range of spatial correlations captured by these methods is limited by the size of the cluster.
For this reason, long-range collective fluctuations are usually described by various diagrammatic extensions of DMFT~\cite{RevModPhys.90.025003}.
Some of these approaches, such as the $GW$+DMFT~\cite{PhysRevLett.92.196402, PhysRevLett.90.086402, PhysRevLett.109.226401, PhysRevB.87.125149, PhysRevB.90.195114, PhysRevB.94.201106, PhysRevB.95.245130}, the triply irreducible
local expansion (TRILEX)~\cite{PhysRevB.92.115109, PhysRevB.93.235124, PhysRevLett.119.166401}, and the dynamical vertex approximation (D$\Gamma$A)~\cite{PhysRevB.75.045118, PhysRevB.80.075104}, as well as most applications of the dual fermion (DF)~\cite{PhysRevB.77.033101, PhysRevB.79.045133, PhysRevLett.102.206401, BRENER2020168310} and the dual boson (DB)~\cite{Rubtsov20121320, PhysRevB.90.235135, PhysRevB.93.045107, PhysRevB.94.205110, PhysRevB.100.165128} theories, take into account only a particular subset of diagrams corresponding to certain channels of instability. 
Others are based on the exact diagrammatic Monte Carlo (DiagMC) method~\cite{PhysRevLett.81.2514, Kozik_2010}, which allows to consider all diagrammatic contributions~\cite{PhysRevB.94.035102, PhysRevB.96.035152, PhysRevB.102.195109}.

$GW$+DMFT is a simple method that is widely used for calculating properties of realistic materials~\cite{PhysRevLett.113.266403, Tomczak_2012, Taranto13, Sakuma13, Tomczak14}.
However, among various long-range fluctuations this approach considers only collective charge excitations and does not account for vertex corrections.
The latter are important for an accurate description of magnetic, optical and transport properties of the system~\cite{Aryasetiawan08, Sponza17, doi:10.1143/JPSJ.75.013703, PhysRevB.80.161105, Katsnelson_2010, PhysRevB.84.085128, Ado_2015, PhysRevLett.117.046601, PhysRevLett.123.036601, PhysRevLett.124.047401, 2020arXiv201009052S}.
More elaborate theories like DF, DB, and D$\Gamma$A, which address all leading collective fluctuations on equal footing, account for vertex corrections and appear to be in a good agreement with numerically exact methods~\cite{PhysRevB.94.035102, PhysRevB.96.035152, PhysRevB.97.125114, PhysRevX.11.011058, PhysRevB.102.195109}.
However, the use of the renormilized local four-point vertex makes all these methods numerically expensive for application to realistic materials~\cite{PhysRevB.95.115107, doi:10.7566/JPSJ.87.041004, PhysRevB.103.035120}.
At the same time, this vertex cannot be simply neglected, because it represents the screened local interaction between electrons.
Therefore, a consistent description of the long-range collective fluctuations requires a theory that combines the simplicity of the $GW$+DMFT diagrammatic scheme with vertex corrections and the equal-footing description of the leading collective modes provided by more elaborate approaches.    

To resolve this issue, a simple consistent diagrammatic extension of DMFT, dubbed ``dual TRILEX'' ($\text{D-TRILEX}$), has recently been proposed in Ref.~\onlinecite{StepanovHarkov}. 
This method is a derivative of the DB approach and is based on a set of Hubbard-Stratonovich transformations of fermionic and bosonic variables.
This allows to consider local correlation effects exactly within the impurity problem of DMFT, and nonlocal effects perturbatively.
The resulting approach considers all leading collective electronic fluctuations on equal footing without any limitation on the range. 
Unlike the DB method, the $\text{D-TRILEX}$ approach relies on a partially bosonized representation for the renormalized local four-point vertex~\cite{StepanovHarkov} that uncovers explicit contributions of different collective modes.
A single (spin or charge) mode approximation of the vertex can be found in prior works~\cite{PhysRevB.94.205110, Stepanov18, Stepanov19}.
Similar approximations for the four-point vertex have also been discussed in Refs.~\onlinecite{Husemann09, Friederich10, Gunnarsson15, Krien19-4}.
However, only the special form of the partially bosonized approximation introduced in the Ref.~\onlinecite{StepanovHarkov} allows to derive the $\text{D-TRILEX}$ theory that with a low computational complexity comparable to $GW$+DMFT or TRILEX methods reproduces the result of the much more elaborate DB theory even in the strongly-interacting regime.
Additionally, unlike the TRILEX method, the $\text{D-TRILEX}$ approach accounts for vertex corrections for both lattice sites that are involved in nonlocal diagrams for the self-energy and the polarization operator.
For instance, this allows to preserve the correct orbital structure of  considered diagrams~\cite{2020arXiv201003433S}. 
Furthermore, the $\text{D-TRILEX}$ approach does not suffer from the famous Fierz ambiguity problem~\cite{Jaeckel03, Baier04, Jaeckel02}, which plagues many theories that perform a partially bosonized description of collective modes. 

The $\text{D-TRILEX}$ theory was introduced only recently~\cite{StepanovHarkov}, and although it has already been extended to a multi-orbital case~\cite{2020arXiv201003433S}, its limits of applicability have not been studied in details so far.
In this work we address this important question and justify the validity of the theory in a broad range of physical parameters.
To this aim we consider a two-dimensional (2D) Hubbard model on a square lattice and compare the performance of the $\text{D-TRILEX}$ approach with its parental DB method and the numerically exact $\text{DiagMC}$ theory. 
Note that in this case the absence of the nonlocal interaction and the bosonic hybridization function identically reduces the DB theory to the DF approach. 
To evaluate the impact of different collective fluctuations on the self-energy, we exploit a partially bosonized representation of the full local four-point vertex and obtain the exact solution of the dual problem with the $\text{DiagMC@DF}$ method~\cite{PhysRevB.96.035152, PhysRevB.102.195109}, as well as the approximate ladder DF solution of the problem. 
In particular, we explicitly investigate the effect of particle-particle fluctuations that enter the four-point vertex function, since they are believed to be negligibly small at standard fillings~\cite{Pao94}.
As the result, we show that exclusion of the irreducible part and transverse contributions from the four-point vertex function often does not lead to a noticeable change of the result, but significantly reduces costs of numerical calculations.

The paper is organized as follows: the Section II contains a brief derivation of the $\text{D-TRILEX}$ theory presented in Ref.~\onlinecite{StepanovHarkov}. In Section III we compare the $\text{D-TRILEX}$ self-energy with the ladder DF, the DiagMC@DF, and the exact $\text{DiagMC}$ results in a broad range of temperatures and local interactions. Finally, the Section IV is devoted to conclusions.

\section{Theory}
\label{sec:Theory}
In this Section we highlight the key points of the derivation of the $\text{D-TRILEX}$ method. 
We begin with the extended Hubbard model described by the following lattice action
\begin{align}
{\cal S}_{\rm latt} = 
&-\sum_{k,\sigma} c^{*}_{k\sigma} {\cal G}^{-1}_{k\sigma} c^{\phantom{*}}_{k\sigma} + U\sum_{q}n^{*}_{q\uparrow}n^{\phantom{*}}_{q\downarrow} 
+ \sum_{q,\vartheta} \xi^{\vartheta} \Bigg\{\rho^{*\,\vartheta}_{q} V^{\vartheta}_{\qv} \, \rho^{\vartheta}_{q} \Bigg\}
\label{eq:action_latt}
\end{align}
Here, the Grassmann variable $c^{(*)}_{k\sigma}$ with the combined index  $k=\{\kv,\nu\}$ describes the annihilation (creation) of an electron with momentum $\kv$, fermionic Matsubara frequency $\nu$, and spin $\sigma = \{\uparrow, \downarrow\}$. 
${\cal G}_{k\sigma} = [i\nu+\mu - \varepsilon_{\kv}]^{-1}$ is the bare lattice Green's function, where $\varepsilon^{\phantom{*}}_{\kv}$ is the dispersion of electrons, and $\mu$ is the chemical potential. 
$U$ describes the on-site Coulomb interaction between electron densities $n_{q\sigma} = \sum_{k} c^{*}_{k+q, \sigma} c^{\phantom{*}}_{k\sigma}$ that depend on momentum $\qv$ and bosonic Matsubara frequency $\omega$ through the combined index $q=\{\qv,\omega\}$. 
For the sake of generality, we also introduce a nonlocal interaction $V_{\qv}$ in different bosonic channels $\vartheta=\{\varsigma,\text{s}\}$, where        ``$\varsigma$'' denotes the particle-hole channel with density ($\varsigma=\text{d}$) and magnetic ($\varsigma=\text{m} = \{x,y,z\}$) components, and ``s'' labels the particle-particle singlet channel. 
For numerical calculations we restrict ourselves to the Hubbard model and set these nonlocal interactions to zero at the end of the derivation.
To shorten the expression for the action we introduce the prefactor $\xi^{\vartheta}$ that for the particle-hole and particle-particle channels respectively reads $\xi^{\varsigma}=1/2$ and $\xi^{\rm s}=1$. 
Corresponding composite bosonic variables ${\rho^{\vartheta}_{q} = n^{\vartheta}_{q}-\langle{}n^{\vartheta}\rangle}$ are introduced as follows
\begin{align}
n^{\varsigma}_{q} &= \sum_{k,\sigma\sigma'} c^{*}_{k+q, \sigma} \, \sigma^{\varsigma}_{\sigma\sigma'} c^{\phantom{*}}_{k\sigma'}
\label{eq:ndm}\\
n^{\rm s}_{q} &= \frac12\sum_{k,\sigma\sigma'}
c^{\phantom{*}}_{q-k, \overline{\sigma}} \, \sigma^{\,z}_{\sigma\sigma'} c^{\phantom{*}}_{k\sigma'}
\label{eq:ns} \\
n^{*\,\rm s}_{q} &= \frac12\sum_{k,\sigma\sigma'}
c^{*}_{k\sigma} \, \sigma^{\,z}_{\sigma\sigma'} c^{*}_{q-k, \overline{\sigma}^{\,\prime}}
\label{eq:ns*}
\end{align}
where $\sigma^{x,y,z}$ are Pauli matrices in the spin space, $\sigma^{\text{d}}$ is the identity matrix in the same space, and $\overline{\sigma}$ is the opposite spin projection to $\sigma$. 
The variable $n^{*\,\varsigma}_{q}$ can be found from the relation $n^{*\,\varsigma}_{q} = n^{\varsigma}_{-q}$, and is introduced to unify notations. 
Note that in the single-band case considered here the composite bosonic variables for the triplet channel are identically equal to zero.

The $\text{D-TRILEX}$ approach, as well as its parental DB theory, performs a diagrammatic expansion around a reference system~\cite{BRENER2020168310}, which in this particular work is given by the exactly solvable effective local impurity problem of DMFT~\cite{RevModPhys.68.13}
\begin{align}
{\cal S}_{\rm imp} =& -\sum_{\nu,\sigma} c^{*}_{\nu\sigma} \left[ i\nu+\mu-\Delta^{\phantom{*}}_{\nu} \right] c^{\phantom{*}}_{\nu\sigma} + U\sum_{\omega}n^{*}_{\omega\uparrow}n^{\phantom{*}}_{\omega\downarrow} 
\label{eq:actionimp}
\end{align}
Here, we introduce the fermionic hybridization function $\Delta_{\nu}$ that aims to describe the screening effect of ``bath'' electrons that surround the given lattice site, which plays a role of the local impurity.
Note that in the current work we do not consider the bosonic hybridization function, which is usually introduced for the impurity problem of the extended dynamical mean-field theory (EDMFT)~\cite{PhysRevB.52.10295, PhysRevLett.77.3391, PhysRevB.61.5184, PhysRevLett.84.3678, PhysRevB.63.115110}. 
In DMFT the fermionic hybridization is determined self-consistently demanding that the local part of the dressed lattice Green's function $G_{k\sigma}$ is equal to the Green's function $g_{\nu\sigma}$ of the impurity problem~\eqref{eq:actionimp}.
To be consistent, the same hybridization function $\Delta_{\nu}$ has to be subtracted from the remaining part of the lattice action ${{\cal S}_{\rm rem} = {\cal S}_{\rm latt} - \sum_{i} {\cal S}_{\rm imp}}$ so that the original lattice problem~\eqref{eq:action_latt} remains unchanged.

The impurity problem~\eqref{eq:actionimp} can be solved numerically exactly using e.g. continious-time quantum Monte Carlo (CTQMC) solvers~\cite{PhysRevB.72.035122, PhysRevLett.97.076405, PhysRevLett.104.146401, RevModPhys.83.349}.
This allows to obtain not only the single-particle Green's function $g_{\nu\sigma}$ and the corresponding self-energy $\Sigma^{\rm imp}_{\nu\sigma}$, but also the two-particle quantities in all bosonic channels $\vartheta$ of interest. 
The latter include the susceptibility $\chi_{\omega}$, the renormalized interaction $w_{\omega}$, and the polarization operator $\Pi^{\rm imp}_{\omega}$, as well as the renormalized local four-point $\Gamma_{\nu\nu'\omega}$ and three-point $\Lambda^{\hspace{-0.05cm}(*)}_{\nu\omega}$ vertex functions. 
The remaining part of the lattice action ${\cal S}_{\rm rem}$ cannot be taken into account exactly.
Instead, it is treated diagrammatically performing an expansion around the impurity problem~\eqref{eq:actionimp}. 
In a consistent way, this procedure can be carried out with the help of a Hubbard-Stratonovich transformation.
The latter allows to rewrite the ${\cal S}_{\rm rem}$ in terms of new fermionic $f$ and bosonic $\varphi$ fields that are dual to original electronic $c$ and composite $\rho$ variables. 
After that, the impurity problem~\eqref{eq:actionimp} with all original variables can be integrated out, which excludes the possibility of the double counting between ${\cal S}_{\rm imp}$ and ${\cal S}_{\rm rem}$ parts of the lattice problem.
This yields the dual boson action (see Ref.~\onlinecite{StepanovHarkov} and Appendix~\ref{app:DB})
\begin{align}
{\cal \tilde{S}}
= &-\sum_{k,\sigma} f^{*}_{k\sigma}\tilde{\cal G}^{-1}_{k\sigma}f^{\phantom{*}}_{k\sigma} 
- \sum_{q,\vartheta} \xi^{\vartheta} \Bigg\{\varphi^{*\,\vartheta}_{q}
\tilde{\cal W}^{\vartheta\,-1}_{q}
\varphi^{\vartheta}_{q} \Bigg\}
+ \tilde{\cal F}[f,\varphi]
\label{eq:dual_action}
\end{align}
Here, the bare dual fermion ${\tilde{\cal G}_{k\sigma} = \check{G}_{k\sigma} - g_{\nu\sigma}}$
and boson ${\tilde{\cal W}^{\vartheta}_{q} = \check{W}^{\vartheta}_{q} - w^{\vartheta}_{\omega}}$
propagators are given by the difference between corresponding EDMFT and impurity quantities. 
To prevent misunderstanding, by the EDMFT Green's function $\check{G}_{k\sigma}$ and the renormalized interaction $\check{W}^{\vartheta}_{q}$ we understand the bare lattice Green's function and the bare interaction that are dressed respectively in the local impurity self-energy and polarization operator via Dyson equations
\begin{align}
\check{G}^{-1}_{k\sigma} &= {\cal G}^{-1}_{k\sigma} - \Sigma^{\rm imp}_{\nu\sigma} \\
\left[\check{W}^{\vartheta}_{q}\right]^{-1} &= \left(U^{\vartheta} + V^{\vartheta}_{\qv}\right)^{-1} - \Pi^{\vartheta\,{\rm imp}}_{\omega}
\label{eq:W_EDMFT}
\end{align}
In this way, bare dual quantities that describe spatial fluctuations already take into account the effect of local correlations.
Note that in the dual problem~\eqref{eq:dual_action} the bare local interaction $U^{\vartheta}$ is introduced as a fictitious quantity that does not affect the result for physical observables (see Appendix~\ref{app:DB}). 
This directly follows from the fact that the DB theory is free from the Fierz ambiguity in decoupling of the local Coulomb interaction $U$ into different channels (see e.g. Ref.~\onlinecite{StepanovHarkov}). 

For actual numerical calculations, the dual interaction $\tilde{\cal F}[f,\varphi]$ is truncated at the two-particle level, which contains only the four-point $\Gamma_{\nu\nu'\omega}$ and three-point $\Lambda^{\hspace{-0.05cm}(*)}_{\nu\omega}$ vertices of the impurity problem~\eqref{eq:actionimp}. 
These quantities are explicitly defined in Appendix~\ref{app:DB}.
With this approximation the theory shows a good agreement with the exact benchmark results~\cite{PhysRevB.94.035102, PhysRevB.96.035152, PhysRevB.97.125114, PhysRevX.11.011058, PhysRevB.102.195109}.
However, it still remains relatively complex due to the presence of the four-point vertex function $\Gamma_{\nu\nu'\omega}$.
The latter depends on three frequencies, so calculating and using it in realistic multi-orbital simulations, which involve the inversion of the Bethe-Salpeter equation in the frequency-orbital space, becomes time consuming numerically~\cite{PhysRevB.95.115107, doi:10.7566/JPSJ.87.041004, PhysRevB.103.035120}. 
To cope with this problem, one can make use of yet another Hubbard-Stratonovich transformation over bosonic variables ${\varphi \xrightarrow{} b}$ that generates an effective four-point interaction in a partially bosonized form
\begin{align}
\Gamma^{\rm d}_{\nu\nu'\omega} &\simeq 
2M^{\rm d}_{\nu\nu'\omega} 
- M^{\rm d}_{\nu,\nu+\omega,\nu'-\nu}
- 3M^{\rm m}_{\nu,\nu+\omega,\nu'-\nu}
+ M^{\rm s}_{\nu,\nu',\omega+\nu+\nu'} \notag\\
\Gamma^{\rm m}_{\nu\nu'\omega} &\simeq
2M^{\rm m}_{\nu\nu'\omega}
+ M^{\rm m}_{\nu,\nu+\omega,\nu'-\nu}
- M^{\rm d}_{\nu,\nu+\omega,\nu'-\nu}
- M^{\rm s}_{\nu,\nu',\omega+\nu+\nu'} \notag\\
\Gamma^{\rm s}_{\nu\nu'\omega} &\simeq 
M^{\rm s}_{\nu\nu'\omega} 
+ \frac12 \left(M^{\rm d}_{\nu,\nu',\omega-\nu-\nu'}
+ M^{\rm d}_{\nu,\omega-\nu',\nu'-\nu}\right) \notag\\
&\hspace{1.23cm}- \frac32 \left(M^{\rm m}_{\nu,\nu',\omega-\nu-\nu'}
+ M^{\rm m}_{\nu,\omega-\nu',\nu'-\nu}\right)
\label{eq:Gamma}
\end{align}
This approximation uncovers the underlying structure of the vertex, which consists of all possible collective electronic fluctuations
\begin{align}
M^{\vartheta}_{\nu\nu'\omega}=\Lambda^{\hspace{-0.05cm}\vartheta}_{\nu\omega} \, \bar{w}^{\vartheta}_{\omega} \, \Lambda^{\hspace{-0.05cm}*\,\vartheta}_{\nu'\omega}
\label{eq:M}
\end{align}
that behave as bosonic modes
\begin{align}
\bar{w}^{\varsigma}_{\omega} &= w^{\varsigma}_{\omega}-U^{\varsigma}/2
\label{eq:w_dm}\\
\bar{w}^{\rm s} &= w^{\rm s}_{\omega}-U^{\rm s}
\label{eq:w_s}
\end{align}

As Ref.~\onlinecite{StepanovHarkov} shows, the partially bosonized representation~\eqref{eq:Gamma} for the four-point vertex can be fine-tuned in such a way that it nearly cancels the exact four-point vertex from the dual action.
Indeed, this approximation does not take into account contributions to the vertex function that cannot be reduced to a single boson propagator.
However, this irreducible part can be completely excluded on the level of the ladder approximation for the vertex by a special choice of bare local interactions in different channels ${U^{\rm d/m}=\pm{}U/2}$ and ${U^{\rm s}=U}$. 
The precise effect of nonladder irreducible contributions on the electronic self-energy is investigated below. 
As a consequence, this specific choice for the bare interactions $U^{\vartheta}$ provides the most accurate partially bosonized approximation for the four-point vertex function given by the Eq.~\eqref{eq:Gamma}.
In addition, it also leads to a correct high-frequency ($\nu\to\infty$ or $\omega\to\infty$) asymptotic behavior of the three-point vertex $\Lambda^{\hspace{-0.05cm}(*)}_{\nu\omega}\to1$~\cite{StepanovHarkov}.

At the same time, it should be noted that this special choice of the bare interaction $U^{\vartheta}$ cannot be obtained by any decoupling of the local Coulomb interaction $U$ into different bosonic channels~\cite{StepanovHarkov}. 
Therefore, it results in a double counting of $U$ in the vertex function, which is explicitly subtracted from the renormalized interaction $w^{\vartheta}_{\omega}$ of the impurity problem in Eqs.~\eqref{eq:w_dm} and~\eqref{eq:w_s}. 
As follows from these equations and the fact that ${w^{\theta}(\omega\to\infty)=U^{\theta}}$, we prefer to keep the bare Coulomb interaction $U$ only in the particle-hole channel.
In this way, it does not contribute to the renormalized singlet interaction $\bar{w}^{\rm s}_{\omega}$~\eqref{eq:w_s}, and becomes equally distributed between density and magnetic channels. 
The reason for such decomposition lies in the fact that the renormalization of the bare interaction in the particle-particle channel is believed to be negligibly small at standard fillings~\cite{Pao94}.
Therefore, the corresponding singlet contribution $M^{\rm s}$, which in the considered form~\eqref{eq:w_s} does not contain the bare interaction $U^{\rm s}$, can be excluded from the theory, as it was consciously done in the previous work~\cite{StepanovHarkov}.
To clarify this statement, we explicitly introduce and investigate the effect of singlet terms in the current work.
Note that although the bare interaction is (partially or fully) subtracted from the local renormalized interactions~\eqref{eq:w_dm} and~\eqref{eq:w_s}, the partially bosonized approximation~\eqref{eq:Gamma} for the four-point vertex has a correct asymptotic behavior at high frequencies $\Gamma^{\,\rm d/m}\to\pm{}U$ and $\Gamma^{\,\rm s}\to{}U$. 
This follows from the fact that the four-point function does not depend on the way how the on-site Coulomb interaction $U$ is distributed between different bosonic channels~\cite{StepanovHarkov}.
For example, although the singlet bosonic mode $\bar{w}^{\rm s}$ does not contain the constant contribution $U^{\rm s}$, the latter is still present in the singlet vertex function $\Gamma^{\rm s}$~\eqref{eq:Gamma} due to transverse charge and spin fluctuations. 

After the last Hubbard-Stratonovich transformation the dual problem~\eqref{eq:dual_action} reduces to a simple action of a partially bosonized dual theory (PBDT) written in terms of fermion $f$ and boson $b$ variables, and the local three-point interaction vertex $\Lambda^{\hspace{-0.05cm}(*)}_{\nu\omega}$ only~\cite{StepanovHarkov}
\begin{align}
{\cal S}_{pb} = 
&-\sum_{k,\sigma} f^{*}_{k\sigma}\tilde{\cal G}^{-1}_{k\sigma}f^{\phantom{*}}_{k\sigma} 
-\sum_{q,\vartheta} \xi^{\vartheta} \Bigg\{ b^{*\,\vartheta}_{q}{\cal W}^{\vartheta \, -1}_{q}b^{\vartheta}_{q} \Bigg\} \notag\\
&+ \sum_{q,k,\vartheta} \xi^{\vartheta} \Bigg\{ \Lambda^{\hspace{-0.05cm}\vartheta}_{\nu\omega} \eta^{*\,\vartheta}_{q,k}
b^{\vartheta}_{q} + \Lambda^{\hspace{-0.05cm}*\,\vartheta}_{\nu\omega} b^{*\,\vartheta}_{q} \eta^{\vartheta}_{q,k} \Bigg\}
\label{eq:fbaction}
\end{align}
where, similarly to Eqs.~\eqref{eq:ndm},~\eqref{eq:ns} and~\eqref{eq:ns*}, we define
\begin{align}
\eta^{\varsigma}_{q,k} &= \sum_{\sigma\sigma'} f^{*}_{k+q, \sigma} \, \sigma^{\varsigma}_{\sigma\sigma'} f^{\phantom{*}}_{k\sigma'} 
\label{eq:eta_dm}\\
\eta^{\rm s}_{q,k} &= \frac12\sum_{\sigma\sigma'}
f^{\phantom{*}}_{q-k, \overline{\sigma}} \, \sigma^{\,z}_{\sigma\sigma'} f^{\phantom{*}}_{k\sigma'} 
\label{eq:eta_s}\\
\eta^{*\,\rm s}_{q,k} &= \frac12\sum_{\sigma\sigma'}
f^{*}_{k\sigma} \, \sigma^{\,z}_{\sigma\sigma'} f^{*}_{q-k, \overline{\sigma}^{\,\prime}}
\label{eq:eta_sast}
\end{align}
The bare Green's function $\tilde{\cal G}_{k\sigma}$ of this new problem~\eqref{eq:fbaction} remains unchanged, and the bare bosonic propagators become 
\begin{align}
{\cal W}^{\varsigma}_{q} &= \check{W}^{\varsigma}_{q} - U^{\varsigma}/2
\label{eq:Wdm}\\
{\cal W}^{\rm s}_{q} &= \check{W}^{\rm s}_{q} - U^{\rm s}
\label{eq:Ws}
\end{align}
where the same exclusion of the double counting between different bosonic channels as in Eqs.~\eqref{eq:w_dm} and~\eqref{eq:w_s} takes place. 

\begin{figure*}[t!]
\includegraphics[width=0.88\linewidth]{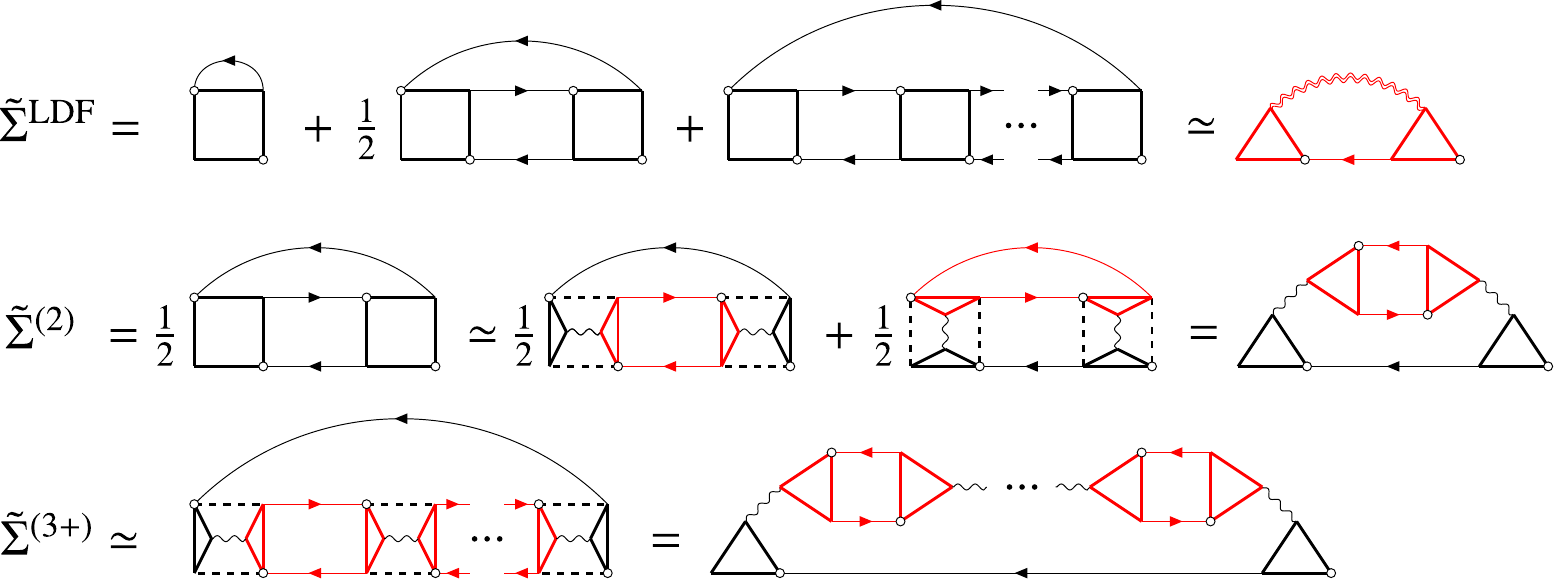}
\caption{\label{fig:diagrams} Top row shows the ladder dual fermion self-energy $\tilde{\Sigma}^{\rm LDF}$, which consists of the first-order Hartree-like term, second-order diagram $\tilde{\Sigma}^{(2)}$, and the rest of the ladder $\tilde{\Sigma}^{(3+)}$. Keeping only longitudinal modes of the partially bosonized representation for the four-point vertex $\Gamma$ (black squares), the LDF self-energy reduces to the self-energy of the $\text{D-TRILEX}$ approach (red diagram at the end of the top row). For $\tilde{\Sigma}^{(2)}$ and $\tilde{\Sigma}^{(3+)}$ the result of this approximation is explicitly shown in the middle and bottom rows, respectively. Red parts of these diagrams that consist of two triangles (three-point vertices) connected by two solid lines (dual Green's functions) is the polarization operator of the $\text{D-TRILEX}$ theory. Wiggly line corresponds to the renormalized interaction.}
\end{figure*}

The simplest set of diagrams for the self-energy and polarization operator used in the $\text{D-TRILEX}$ approach~\cite{StepanovHarkov}
\begin{align}
&\tilde{\Sigma}_{k\sigma} 
= -\sum_{q,\varsigma} \Bigg\{ \Lambda^{\hspace{-0.05cm}\varsigma}_{\nu\omega} \tilde{G}_{q+k,\sigma} W^{\varsigma}_{q} \Lambda^{\hspace{-0.05cm}*\,\varsigma}_{\nu\omega} 
- \Lambda^{\hspace{-0.05cm}\rm s}_{\nu\omega} \tilde{G}_{q-k,\overline{\sigma}} W^{\rm s}_{q} \Lambda^{\hspace{-0.05cm}*\,\rm s}_{\nu\omega} \Bigg\}
\label{eq:Sigma_dual}\\
&\hspace{1.8cm}\bar{\Pi}^{\varsigma}_{q}
= +\sum_{k,\sigma} \Lambda^{\hspace{-0.05cm}*\,\varsigma}_{\nu\omega} \tilde{G}_{k\sigma} \tilde{G}_{q+k,\sigma} \Lambda^{\hspace{-0.05cm}\varsigma}_{\nu\omega} \label{eq:Pi_dual} \\
&\hspace{1.8cm}\bar{\Pi}^{\rm s}_{q} 
= -\sum_{k} \Lambda^{\hspace{-0.05cm}*\,\rm s}_{\nu\omega} \tilde{G}_{k\uparrow} \tilde{G}_{q-k\downarrow} \Lambda^{\hspace{-0.05cm}\rm s}_{\nu\omega} \label{eq:Pi_dual_s}
\end{align}
can be obtained from the analog of the Almbladh functional~\cite{doi:10.1142/S0217979299000436} ${\Psi[\tilde{G}, W, \Lambda] = \frac12 \tilde{G} \Lambda^{\hspace{-0.05cm}\vartheta} W^{\vartheta} \Lambda^{\hspace{-0.05cm}*\vartheta} \tilde{G}}$ introduced in the dual space. This ensures the consistency between single- and two-particle quantities produced by the theory. 
Here, $\tilde{G}_{k\sigma}$ and $W^{\vartheta}_{q}$ are full propagators of the fermion-boson problem~\eqref{eq:fbaction} given by Dyson equations
\begin{align}
\tilde{G}_{k\sigma}^{-1} &= \tilde{\cal G}_{k\sigma}^{-1} -  \tilde{\Sigma}_{k\sigma}
\label{eq:Dressed_G}\\
W_{q}^{\vartheta~-1} &= {\cal W}_{q}^{\vartheta~-1} - \bar{\Pi}^{\vartheta}_{q}
\label{eq:Dressed_W}
\end{align}
This simple $GW$-like diagrammatics~\eqref{eq:Sigma_dual},~\eqref{eq:Pi_dual}, and~\eqref{eq:Pi_dual_s} of the $\text{D-TRILEX}$ approach can also be related to its parental DB theory. 
For simplicity, the main text of the paper contains only a sketch of this derivation presented in Fig.~\ref{fig:diagrams}.
Also, here we only consider the case when the nonlocal interaction is discarded ($V^{\vartheta}_{\qv}=0$).
Then, the DB theory~\eqref{eq:dual_action} identically coincides with the DF approach, and the dual self-energy in the ladder approximation takes the form of $\tilde{\Sigma}^{\rm LDF}$~\cite{PhysRevLett.102.206401} displayed in the first line of Fig.~\ref{fig:diagrams}.
This expression can be obtained using the Schwinger-Dyson equation for the dual self-energy~\cite{hafermann2010numerical}.
As the result, the second-order contribution $\tilde{\Sigma}^{(2)}$ has a ``1/2'' prefactor that does not appear for the rest of the ladder self-energy $\tilde{\Sigma}^{(3+)}$~\cite{PhysRevB.79.045133}.
As two subsequent lines in Fig.~\ref{fig:diagrams} show, if one uses the partially bosonized representation~\eqref{eq:Gamma} for every vertex function that enters $\tilde{\Sigma}^{\rm LDF}$ and keeps only longitudinal contributions in this approximation, the dual self-energy immediately reduces to the $\text{D-TRILEX}$ form~\eqref{eq:Sigma_dual}.
By longitudinal contributions we understand $M^{\vartheta}_{\nu\nu'\omega}$ terms, where the bosonic propagator $\bar{w}^{\vartheta}_{\omega}$ carries the main bosonic frequency $\omega$. 
The exclusion of transverse particle-hole and particle-particle fluctuations from the four-point vertex can be motivated by the fact that their contributions to the self-energy partially cancel each-other, which is demonstrated in the Section~\ref{sec:Results_B}. 
The explicit analytical derivation of the relation between $\text{D-TRILEX}$ and ladder DB (LDB) self-energies for the general case when the nonlocal interaction is not neglected is shown in Appendix~\ref{app:Sigma}.
This result demonstrates the important advantage of the $\text{D-TRILEX}$ theory over its parental LDB method, which drastically reduces costs of numerical calculations.
Thus, although the $\text{D-TRILEX}$ approach accounts for the main longitudinal part of the full two-particle ladder fluctuation, the calculation of the self-energy~\eqref{eq:Sigma_dual} and polarization operators~\eqref{eq:Pi_dual} and~\eqref{eq:Pi_dual_s} does not require the inversion of the Bethe-Salpeter equation in the momentum-frequency space.
In the ladder DF/DB theory the inversion of the Bethe-Salpeter equation in the frequency space cannot be avoided due to a three-frequency dependence of the local vertex function $\Gamma_{\nu\nu'\omega}$.

We note that the diagrammatic expansion in DF, DB and $\text{D-TRILEX}$ methods is performed in the dual space. 
The self-energy of the original lattice problem~\eqref{eq:action_latt} can be obtained from the following exact relation~\cite{Rubtsov20121320, PhysRevB.93.045107, PhysRevB.94.205110} 
\begin{align}
\Sigma^{\text{latt}}_{k\sigma} = \Sigma^{\text{imp}}_{\omega\sigma} + \frac{\tilde{\Sigma}_{k\sigma}}{1+g_{\nu\sigma}\tilde{\Sigma}_{k\sigma}}
\label{eq:self-en}
\end{align}
Here, the dual contribution to the lattice self-energy comes with the denominator $(1+g_{\nu\sigma}\tilde{\Sigma}_{k\sigma})$ that excludes unphysical terms from the Dyson equation for the lattice Green's function~\cite{BRENER2020168310}. 
Note that the expression~\eqref{eq:self-en} requires the explicit calculation of the impurity self-energy that cannot be measured directly as a single-particle correlation function.
Instead, $\Sigma^{\rm imp}_{\nu\sigma}$ is usually obtained by inverting the Dyson equation for the full impurity Green's function $g_{\nu\sigma}$, which makes the result noisy at high frequencies.
The noise in the self-energy can be reduced using the improved estimators method that, however, requires the measurement of higher-order correlations functions~\cite{Hafermann12, Gunacker16, PhysRevB.100.075119}.
Therefore, for calculation of the lattice Green's function it is more convenient to use another exact relation that does not involve $\Sigma^{\rm imp}_{\nu\sigma}$~\cite{Krivenko2010, Rubtsov20121320} 
\begin{align}
G_{k\sigma}^{-1} = \left[g_{\nu\sigma} + g_{\nu\sigma} \tilde{\Sigma}_{k\sigma}g_{\nu\sigma}\right]^{-1} + \Delta_{\nu} - \varepsilon_{\kv}
\end{align}
This expression for the dressed lattice Green's function $G_{k\sigma}$ completes the derivation of the $\text{D-TRILEX}$ approach.

\section{Results}

\subsection{Comparing different methods}
\label{sec:Results_A}

\begin{table}[b!]
\caption{\label{table}Summary of considered methods that specifies the form of the local four-point vertex and types of diagrams that are used for the diagrammatic expansion. The square corresponds to the exact four-point vertex $\Gamma_{\nu\nu'\omega}$ of the impurity problem. Longitudinal/transverse components $M_{\nu\nu'\omega}$ that enter the partially bosonized approximation for the vertex~\eqref{eq:Gamma} are depicted by two triangles connected by the wavy line placed horizontally/vertically. The ``$-$'' sign in front of the transverse contribution 
corresponds to the antisymmetric form of the vertex~\eqref{eq:Gamma}. The $\text{DiagMC@PBDT-s}$ approximation can be obtained from the $\text{DiagMC@PBDT}$ method by excluding the singlet contribution $M^{\rm s}_{\nu,\nu',\omega+\nu+\nu'}$ from the partially bozonized vertex~\eqref{eq:Gamma}.}
\centering
\begin{tabular}{ |c|c|c| }
 \hline
 ~Method~ & ~Four-point vertex~ & ~Types of diagrams~ \\ 
 \hline
 ~D-TRILEX~ & $\vcenter{\hbox{
\scalebox{.4}{ 
\begin{tikzpicture}
\begin{feynman}[small]
    
    \vertex (z1);
    \vertex[below=0.25cm of z1] (b1);
    \vertex[right=0.5cm of b1] (a1);
    \vertex[below=0.5cm of a1] (a2);
    \vertex[below=1.0cm of b1] (a3);
    [edges=fermion]
    \vertex[right=0.5cm of a2] (c1);
    \vertex[above=0.5cm of c1] (c2);
    \vertex[right=0.5cm of c2] (c3);
    \vertex[below=1.0cm of c3] (c4);
    \diagram* [very thick]{
      {
        (b1) -- (a2) -- (a3) -- (b1),
      },
      };
    \draw[decorate,decoration={snake,segment length=0.18cm,amplitude=0.15cm},thick](a2) -- (c1);
    \diagram* [very thick]{
      {
        (c1) -- (c3) -- (c4) -- (c1),
      },
      };
    \draw[white] (z1) circle(0.65mm);
\end{feynman}
\end{tikzpicture}
 }
 }}$ & ladder \\ 
 LDF & $\vcenter{\hbox{
\scalebox{.4}{ 
\begin{tikzpicture}
\begin{feynman}[small]
    \vertex (z1);
    \vertex[below=0.25cm of z1] (b1);
    \vertex[right=2cm of b1] (g1);

    \vertex[left=1.2cm of b1] (b-1) ;
    \vertex[right=2.5cm of b1] (b-2) ;
    \vertex[below=0.55cm of b-1] (b00);
    
    \vertex[right=1cm of b1] (g2);
    \vertex[below=1cm of g2] (g3);
    \vertex[below=1cm of b1] (g4);
    
    \diagram* [very thick]{
      {
        (b1) -- (g2) -- (g3) -- (g4) -- (b1),
      },
      };
    
\end{feynman}
\end{tikzpicture}
 }
 }}$ & ladder \\ 
 $\text{DiagMC@DF}$ & $\vcenter{\hbox{
\scalebox{.4}{ 
\begin{tikzpicture}
\begin{feynman}[small]
    \vertex (z1);
    \vertex[below=0.25cm of z1] (b1);
    \vertex[right=2cm of b1] (g1);

    \vertex[left=1.2cm of b1] (b-1) ;
    \vertex[right=2.5cm of b1] (b-2) ;
    \vertex[below=0.55cm of b-1] (b00);
    
    \vertex[right=1cm of b1] (g2);
    \vertex[below=1cm of g2] (g3);
    \vertex[below=1cm of b1] (g4);
    
    \diagram* [very thick]{
      {
        (b1) -- (g2) -- (g3) -- (g4) -- (b1),
      },
      };
    
    \draw[white] (z1) circle(0.65mm);
\end{feynman}
\end{tikzpicture}
 }
 }}$ & all \\
 ~DiagMC@PBDT~ & $\vcenter{\hbox{
\scalebox{.4}{ 
\begin{tikzpicture}
\begin{feynman}[small]
    \vertex (z1);
    \vertex[below=2.0cm of z1] (z2);
    \vertex[below=0.5cm of z1] (b1);
    \vertex[right=0.5cm of b1] (a1);
    \vertex[below=0.5cm of a1] (a2);
    \vertex[below=1.0cm of b1] (a3);
    [edges=fermion]
    \vertex[right=0.5cm of a2] (c1);
    \vertex[above=0.5cm of c1] (c2);
    \vertex[right=0.5cm of c2] (c3);
    \vertex[below=1.0cm of c3] (c4);
    
    \diagram* [very thick]{
      {
        (b1) -- (a2) -- (a3) -- (b1),
      },
      };
    \draw[decorate,decoration={snake,segment length=0.18cm,amplitude=0.15cm},thick](a2) -- (c1);
    \diagram* [very thick]{
      {
        (c1) -- (c3) -- (c4) -- (c1),
      },
      };
    \diagram* [very thick]{
      {
        (c1) -- (c3) -- (c4) -- (c1),
      },
      };  
      
    \vertex[right=0.75cm of c1] (d1)
    {\huge{$-$}};
    \vertex[right=0.5cm of d1] (e1);
    \vertex[above=0.75cm of e1] (e2);
    \vertex[right=1.0cm of e2] (e3);
    \vertex[below=0.5cm of e3] (e4);
    \vertex[left=0.5cm of e4] (e5);
    
    \vertex[below=0.5cm of e5] (f1);
    \vertex[right=0.5cm of f1] (f2);
    \vertex[below=0.5cm of f2] (f3);
    \vertex[left=1.0cm of f3] (f4);
    
    \diagram* [very thick]{
      {
        (e2) -- (e3) -- (e5) -- (e2),
      },
      };
      \draw[decorate,decoration={snake,segment length=0.18cm,amplitude=0.15cm},thick](e5) -- (f1);
      \diagram* [very thick]{
      {
        (f1) -- (f3) -- (f4) -- (f1),
      },
      };
    
    \draw[white] (z1) circle(0.65mm);
    \draw[white] (z2) circle(0.65mm);
\end{feynman}
\end{tikzpicture}
 }
 }}$ & all \\
 \hline
\end{tabular}
\end{table}

\begin{figure*}[t!]
\includegraphics[width=0.95\linewidth]{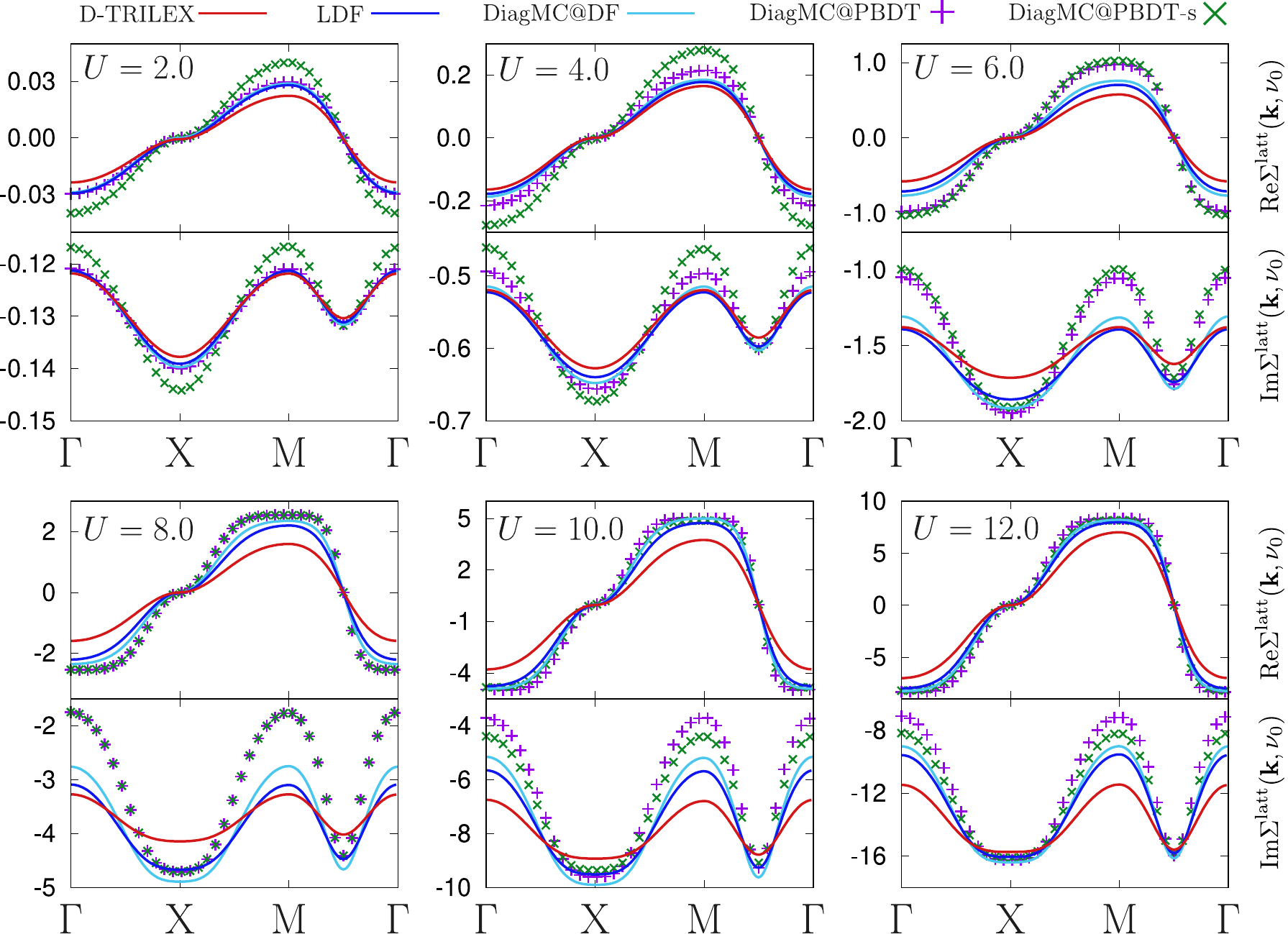}
\caption{\label{fig:compare_sigma} The lattice self-energy $\Sigma^{\rm latt}_{{\bf k},\nu}$ obtained for the first Matsubara frequency $\nu_0=\pi/\beta$ for the inverse temperature $\beta=2$ along the high symmetry path in momentum space ${\bf k}$. The value of the on-site Coulomb potential $U$ for which the calculation was performed is specified in panels. Upper and lower part of each panel corresponds to real and imaginary part of the self energy, respectively. Results are obtained using $\text{D-TRILEX}$ (red line), LDF (dark blue line), $\text{DiagMC@DF}$ (light blue line), DiagMC@PBDB (purple crosses), and DiagMC@PBDB-s (green crosses) methods.}
\end{figure*}

In this section we consistently investigate the effect of different contributions that make up the renormalized local four-point vertex on the electronic self-energy. 
To this aim we consider a 2D Hubbard model on a square lattice with the neareast-neighbor hopping amplitude $t=1$ and different values of the on-site Coulomb potential $U$. Numerical calculations are performed at half-filling unless the other filling is explicitly specified. 
All nonlocal interactions are set to zero $V^{\vartheta}_{\qv}=0$.
Note that in this case the renormalized interaction of EDMFT $\check{W}^{\vartheta}_{q}$~\eqref{eq:W_EDMFT} coincides with the impurity $w^{\vartheta}_{\omega}$, and the dual boson propagator $\tilde{W}^{\vartheta}_{q}$ becomes zero.
As a consequence, the DB action~\eqref{eq:dual_action} reduces to the DF problem~\cite{PhysRevB.77.033101}.
Restricting the interaction $\tilde{\cal F}[f]$ to the four-point vertex function $\Gamma_{\nu\nu'\omega}$, the dual problem~\eqref{eq:dual_action} can be solved numerically exactly within the diagrammatic Monte Carlo method for dual fermions ($\text{DiagMC@DF}$)~\cite{PhysRevB.96.035152, PhysRevB.94.035102, PhysRevB.102.195109}.
In the current work we perform $\text{DiagMC@DF}$ calculations on the basis of the converged DMFT solution of the lattice problem~\eqref{eq:action_latt}.
The corresponding single-site impurity problem of DMFT~\eqref{eq:actionimp} is solved numerically exactly using the open source CT-HYB solver~\cite{HAFERMANN20131280, PhysRevB.89.235128} based on ALPS libraries~\cite{Bauer_2011}.
After that, the calculated bare dual Green's function $\tilde{\cal G}_{k\sigma}$ and the local four-point vertex function $\Gamma_{\nu\nu'\omega}$ are used as building blocks for a diagrammatic expansion.
A detailed description of the $\text{DiagMC@DF}$ method can be found in Ref.~\onlinecite{PhysRevB.96.035152}. 
For physical parameters considered in this work, the converged result for the dual self-energy is achieved at the fifth order of expansion.

Different levels of approximation for the four-point vertex can be investigated to reveal the effect of contributions that are not accounted for in the $\text{D-TRILEX}$ theory.
In particular, the contribution to the self-energy that stems from the irreducible part of the four-point vertex function can be identified by comparing the exact $\text{DiagMC@DF}$ solution of the dual problem~\eqref{eq:dual_action} with the result of another $\text{DiagMC}$ calculation, where the exact local vertex $\Gamma_{\nu\nu'\omega}$ is replaced by its partially bosonized approximation~\eqref{eq:Gamma}. 
Hereinafter this method is referred to as $\text{DiagMC@PBDT}$ and corresponds to the exact evaluation of the self-energy of the partially bosonized dual theory~\eqref{eq:fbaction}.
The next level of approximation that allows to observe the effect of collective fluctuations in the singlet channel can be achieved by performing $\text{DiagMC}$ calculations with the partially bosonized vertex~\eqref{eq:Gamma} where all $M^{\rm s}$ terms are neglected. 
This calculation is referred to as \text{DiagMC@PBDT-s}.

We note that the $\text{DiagMC@DF}$ method does not distinguish between longitudinal and transverse components of the four-point vertex, because the Monte Carlo sampling considers all possible topologies of diagrams. 
As has been discussed in the Section~\ref{sec:Theory}, the contribution of these modes can be disentangled comparing the self-energy of the ladder dual fermion (LDF) approach, which exploits the exact local four-point vertex, with the result of the $\text{D-TRILEX}$ theory, where only longitudinal modes are taken into account.
These calculations are also performed on the basis of the converged DMFT solution, so that the local impurity problem remains the same for all compared theories. 
Note that the LDF and $\text{D-TRILEX}$ results for the self-energy are obtained within the self-consistent scheme in terms of dressed fermionic~\eqref{eq:Dressed_G} and bosonic~\eqref{eq:Dressed_W} propagators.
When possible, we compare our results with the exact $\text{DiagMC}$ solution~\cite{PhysRevLett.119.045701, PhysRevB.97.085117, PhysRevB.100.121102} of the lattice problem~\eqref{eq:action_latt} that was kindly provided by the authors of Refs.~\onlinecite{PhysRevX.11.011058, Wu17}. All methods are summarized in Table~\ref{table}.

\subsection{Effect of the local interaction}
\label{sec:Results_B}

First, we make a scan over a broad range of local Coulomb interactions $U$ at a fixed temperature.
We note that in two dimensions DMFT predicts the N\'eel transition at a finite temperature.
This transition is forbidden by Mermin-Wagner theorem~\cite{PhysRevLett.17.1133} and thus is an artefact of the DMFT theory.
However, since the $\text{DiagMC@DF}$ method uses the DMFT impurity problem as a staring point for the diagrammatic expansion, the $\text{DiagMC@DF}$ theory shows difficult convergence or even divergent result close to the DMFT N\'eel point~\cite{PhysRevB.94.035102, PhysRevB.96.035152}.
For this reason, calculations are performed at the inverse temperature $\beta=2$, so that the $\text{DiagMC}$ results are not affected by any convergence issue.
Fig.~\ref{fig:compare_sigma} shows the lattice self-energy~\eqref{eq:self-en} calculated for all above-mentioned approaches (see Section~\ref{sec:Results_A}).
Note that the self-energy does not contain the constant Hartree part that is equal to $U/2$ at half-filling.
The results are obtained for the first Matsubara frequency $\nu_0=\pi/\beta$ along the high symmetry path that connects $\Gamma=(0,0)$, ${\rm X}=(0, \pi)$, and ${\rm M}=(\pi, \pi)$ points in momentum space ${\bf k}=(k_{x}, k_{y})$.

Let us first consider the effect of the irreducible part of the four-point vertex comparing the self-energy of $\text{DiagMC@DF}$ (light blue line) and $\text{DiagMC@PBDT}$ (purple crosses) approaches shown in Fig.~\ref{fig:compare_sigma}.
We find that at $U=2$ both methods produce identical results, which means that in a weakly-correlated regime the irreducible contributions to the vertex do not affect the self-energy.
Upon increasing the local interaction, the discrepancy between these two methods also increases and is noticeable the most in the strongly-correlated regime at $U=8$, which is equal to the bandwidth.
After that, at very large interactions $U=10$ and $U=12$ the real part of the $\text{DiagMC@PBDT}$ self-energy again nearly coincides with the one of the $\text{DiagMC@DF}$ approach.
The agreement in the imaginary part of the self-energy also improves, but the discrepancy between these two methods remains noticeable.

\begin{figure}[t!]
\includegraphics[width=1\linewidth]{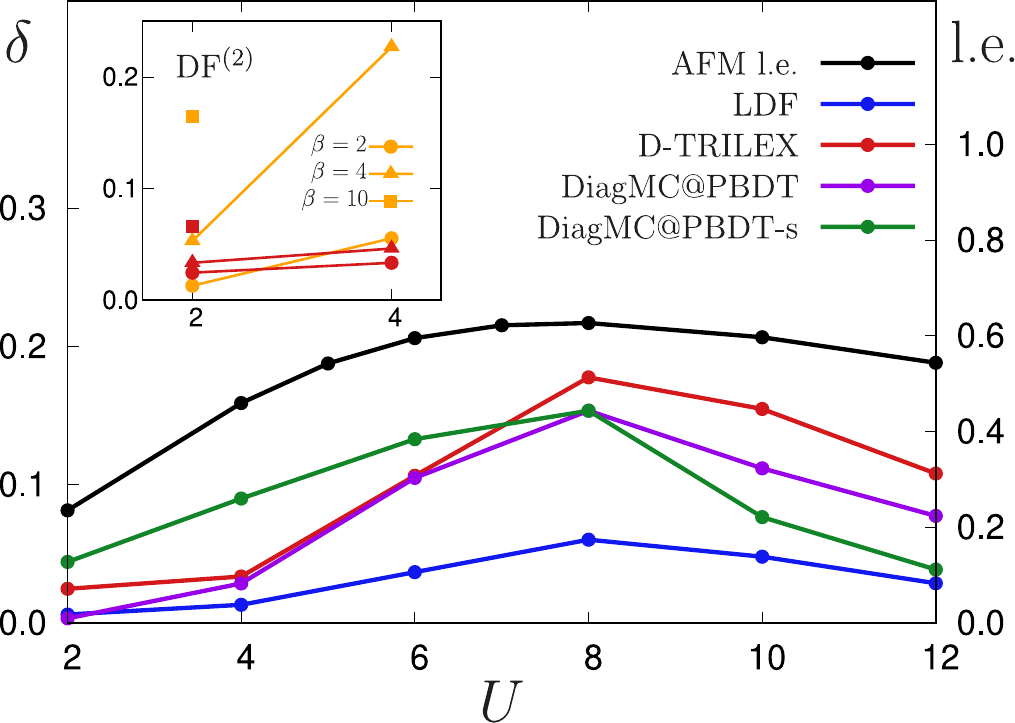}
\caption{\label{fig:Ev_and_deviation} The normalized deviation $\delta$ calculated for LDF (blue), $\text{D-TRILEX}$ (red), $\text{DiagMC@PBDT}$ (purple), and $\text{DiagMC@PBDT-s}$ (green) approximations with respect to the reference Diag@DF result. The black line shows the leading eigenvalue (l.e.) of AFM fluctuations. The vertical left axis shows the scale for the normalized deviation, while the vertical right axis displays values for the leading eigenvalue. The inset compares $\delta$ obtained for $\text{D-TRILEX}$ and second-order DF (DF$^{(2)}$, orange) approaches for different inverse temperatures $\beta=2$ (circles), $\beta=4$ (triangles), and $\beta=10$ (squares).}
\end{figure}

To quantify the difference of the given self-energy from the reference $\text{DiagMC@DF}$ result we calculate the following normalized deviation 
\begin{align}
\delta = \sum_{\bf k}\left|\frac{\Sigma^{\rm ref}_{{\bf k}, \nu_0} - \Sigma^{\phantom{f}}_{{\bf k}, \nu_0}}{\Sigma^{\rm ref}_{{\bf k}, \nu_0}}\right|.
\label{eq:norm_diff}
\end{align}
A similar quantity but for only one k-point was introduced in Ref.~\onlinecite{PhysRevB.102.195109}.
The corresponding result for all considered approaches is presented in Fig.~\ref{fig:Ev_and_deviation}.
We find that the normalized deviation of the $\text{DiagMC@PBDT}$ method reaches its maximum value $\delta = 15\%$ at $U=8$.  
As has been pointed out in the Section~\ref{sec:Theory}, the irreducible part can be excluded from the renormalized four-point vertex only in the ladder approximation.
In the strongly-correlated regime nonladder diagrams become important~\cite{PhysRevB.94.035102, PhysRevB.96.035152, PhysRevB.102.195109}, which is also confirmed by the increase of the normalized deviation of the LDF approach (blue line in Fig.~\ref{fig:Ev_and_deviation}).
Consequently, the contribution of the irreducible part of the vertex to the electronic self-energy also becomes noticeable.  
We would like to emphasise that by the strength of electronic correlations we mean not only the strength of the interaction, but also the proximity of the system to an instability.
The latter can be estimated by the leading eigenvalue (l.e.) of the Bethe-Salpeter equation of the LDF theory~\cite{PhysRevLett.102.206401, Otsuki14} (black line in Fig.~\ref{fig:Ev_and_deviation}), which in our case indicates the strength of antiferromagnetic (AFM) fluctuations.

In the next step we investigate the effect of an additional exclusion of all singlet contributions from the partially bosonized four-point vertex~\eqref{eq:Gamma}.
At small ($U=2$) and moderate ($U=4$) interactions this immediately leads to a large discrepancy between $\text{DiagMC@PBDT-s}$ (green crosses) and $\text{DiagMC@PBDT}$ (purple crosses) results for the self-energy presented in Fig.~\ref{fig:compare_sigma}.
In addition, from Fig.~\ref{fig:Ev_and_deviation} we find that for these values of the interaction the $\text{DiagMC@PBDT-s}$ strongly differs from the reference result, while the $\text{DiagMC@PBDT}$ performs reasonably well.
Therefore, one can conclude that singlet fluctuations play an important role in weakly- and moderately-correlated regime. 
At a first glance this observation is in a contradiction with the statement that particle-particle fluctuations are believed to be negligibly small at standard fillings~\cite{Pao94}.
This point is clarified below when we discuss the result of the $\text{D-TRILEX}$ approach.
Increasing the interaction to $U=6$, makes the discrepancy between $\text{DiagMC@PBDT}$ and $\text{DiagMC@PBDT-s}$ results rapidly decrease, and in the strongly-correlated regime ($U=8$) both methods produce identical results.
Remarkably, for $U=10$ and $U=12$ the $\text{DiagMC@PBDT-s}$ method shows the best agreement with the $\text{DiagMC@DF}$ result among all considered DiagMC-based approximations.
This result suggests that in the regime of very large interactions contributions to the self-energy that stem from the irreducible and singlet parts of the renormalized four-point vertex, which are not considered in the $\text{DiagMC@PBDT-s}$ theory, nearly cancel each other.

Finally, let us consider the $\text{D-TRILEX}$ method that can be obtained from the LDF theory excluding the irreducible part and neglecting transverse particle-hole and particle-particle fluctuations from the exact local impurity four-point vertex.
As Fig.~\ref{fig:compare_sigma} shows, the best agreement between the $\text{D-TRILEX}$ (red line) and the reference $\text{DiagMC@DF}$ (light blue line) results for the imaginary part of the self-energy occurs at $U=2$.
At small and moderate values of $U$, the $\text{D-TRILEX}$ self-energy seems to be pinned to the LDF result (dark blue line) at $\Gamma$ and M points. 
Therefore, the difference between these two methods is mostly visible around local minima located at antinodal ${\text{AN}=(0, \pi)}$ and nodal ${\text{N}=(\pi/2, \pi/2)}$ points.
This difference increases with the interaction, and the observed trend persists up to $U=6$. 
At larger interactions, when the value of the self-energy at local minima becomes similar, the $\text{D-TRILEX}$ result shifts downwards, and at ${U=12}$ becomes pinned to the LDF result at N and AN points.

The discrepancy between the $\text{D-TRILEX}$ and the reference results for the real part of the self-energy also increases with the interaction up to $U=8$, and after that decreases again for very large interaction strengths. 
However, here the best agreement with the exact result is achieved at $U=4$ (see red line in Fig.~\ref{fig:compare_sigma}). 
It can be explained by the fact, that in the perturbative regime of small interactions ($U=2$) and high temperatures ($\beta=2$) the second-order dual self-energy $\tilde{\Sigma}^{(2)}$ gives the main contribution to the nonlocal part of the total self-energy~\cite{PhysRevB.91.235114, PhysRevB.94.035102, PhysRevB.96.035152, PhysRevB.102.195109}.  
The $\text{D-TRILEX}$ theory is not based on a perturbation expansion, because it takes into account only a particular ($GW$-like) subset of diagrams. 
For this reason, this simple theory does not fully reproduce the second-order self-energy $\tilde{\Sigma}^{(2)}$, which leads to a slight underestimation of the result as discussed in Appendix~\ref{app:Sigma}. 
On the contrary, the $\text{D-TRILEX}$ approach correctly accounts for the screening of the interaction that is represented by longitudinal part of the infinite two-particle ladder in all bosonic channels.
At lower temperatures and/or larger interactions, when the system enters the correlated regime, these types of diagrams become more important than the second-order self-energy.
To illustrate this point, we also obtained the normalized deviation for the $\text{D-TRILEX}$ approach for $\beta=4$ (for $U=2$ and $U=4$) and $\beta=10$ (for $U=2$), and compared it with $\delta$ calculated for the second-order DF (DF$^{(2)}$) approximation that considers only $\tilde{\Sigma}^{(2)}$ contribution to the dual self-energy. 
The corresponding result is shown in the inset of Fig.~\ref{fig:Ev_and_deviation}. 
As expected, the accuracy of the DF$^{(2)}$ approximation rapidly decreases with the temperature and becomes $\delta=16.5\%$ (for $\beta=10$ and $U=2$) and $\delta=22.5\%$ (for $\beta=4$ and $U=4$) in the regime, which is yet above the DMFT N\'eel point $\beta_{N}\simeq12.5$ for $U=2$ and $\beta_{N}\simeq4.3$ for $U=4$.
At the same time, the $\text{D-TRILEX}$ theory remains in a reasonable agreement with the reference result.

Fig.~\ref{fig:Ev_and_deviation} shows that in the regime of weak and moderate interactions the $\text{D-TRILEX}$ self-energy is relatively close to the $\text{DiagMC@DF}$ result ($\delta=2\%$ for $U=2$ and $\delta=3\%$ for $U=4$). 
This fact looks paradoxical at a first glance, because the $\text{D-TRILEX}$ method does not take into account singlet fluctuations that were found to be important in this regime of interactions.
To explain this result, let us first note that at $U \leq{} 4$ the LDF method is in a very good agreement with the $\text{DiagMC@DF}$ theory.
Therefore, in the weakly- and moderately-correlated regime ladder diagrams provide the most important contribution to the self-energy.
This fact allows for a direct comparison of the self-energies produced by ladder DF and $\text{D-TRILEX}$ methods with the result of DiagMC@ methods that account for all diagrammatic contributions.
Note however, that all DiagMC-based schemes tend to overestimate the reference result, while ladder-like approaches underestimate it. 
Therefore, the normalized deviation presented in Fig.~\ref{fig:Ev_and_deviation} should be compared cautiously.
Fig.~\ref{fig:compare_sigma} shows that $\text{D-TRILEX}$ and $\text{DiagMC@PBDT}$ self-energies obtained at $U=2$ and $U=4$ are very close to the reference result.
Both methods do not take into account the irreducible part of the four-point vertex function, but the $\text{D-TRILEX}$ approach additionally neglects all transverse particle-hole and particle-particle modes.
Keeping in mind that for these interaction strengths the exclusion of only singlet fluctuations leads to a large overestimation of the self-energy, we can conclude that transverse particle-hole and particle-particle fluctuations partially screen each other. 
This means that the exclusion of both types of vertical insertions in diagrams, as it is done in the $\text{D-TRILEX}$ theory, turns out to be a good approximation in the weakly- and moderately-correlated regime.
On the other hand, excluding only one channel leaves the other channel unscreened, which results in a large contribution to the self-energy. 

Remarkably, the normalized deviation for all considered approximations shown in Fig.~\ref{fig:Ev_and_deviation} resembles the behavior of the leading eigenvalue of the magnetic channel (black line).
For instance, the $\text{D-TRILEX}$ and LDF methods show the largest discrepancy with the $\text{DiagMC@DF}$ result exactly in the region where the l.e. is maximal.
As has been pointed out in Ref.~\onlinecite{PhysRevB.102.224423}, approaching an instability leads to collective fluctuations becoming strongly anharmonic, which cannot be captured by simple diagrammatic theories.
Consequently, in this regime transverse momentum-dependent fluctuations are expected to be important.
At $U=10$ and $U=12$ the agreement of the $\text{D-TRILEX}$ theory with the $\text{DiagMC@DF}$ result improves again. 
Above we have found that at very large interaction strengths contributions to the self-energy that stem from singlet and irreducible parts of the vertex partially cancel each other.
This result suggests that the effect of remaining transverse particle-hole fluctuations becomes weaker at very large interactions, which again justifies the applicability of the $\text{D-TRILEX}$ theory.

\begin{figure}[t!]
\includegraphics[width=1.0\linewidth]{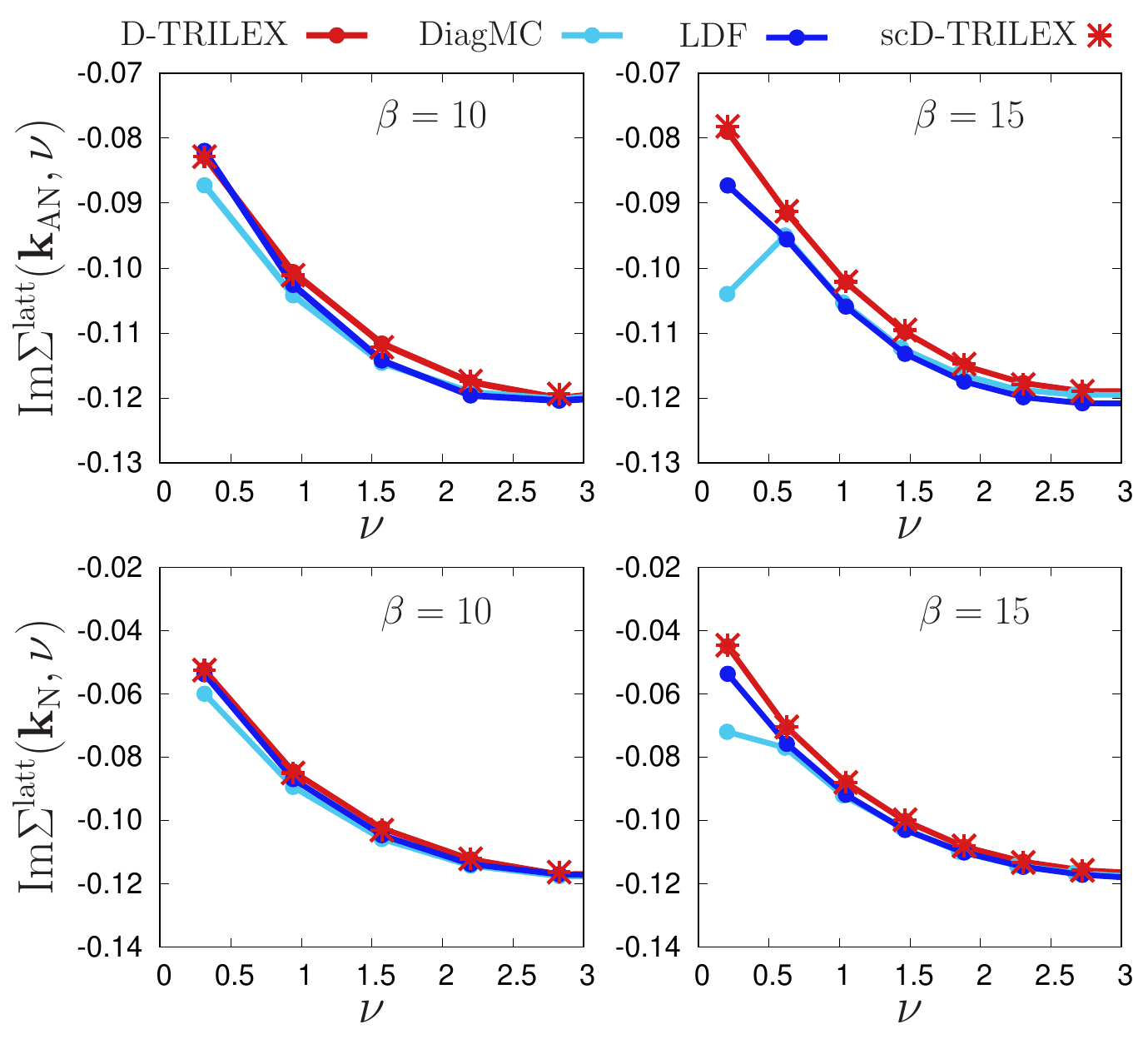}
\caption{\label{fig:ImSigma_U2} Imaginary part of the lattice self-energy as the function of Matsubara frequency $\nu$ obtained for $U=2$ at the antinodal ${\text{AN}=(0, \pi)}$ (top row) and nodal ${\text{N}=(\pi/2, \pi/2)}$ (bottom row) points. 
Results are calculated at the inverse temperature $\beta=10$ (left column) and $\beta=15$ (right column) using $\text{D-TRILEX}$ (red line), $\text{scD-TRILEX}$ (red stars), and LDF (dark blue line) approaches. The $\text{DiagMC}$ results (light blue line) are provided by the authors of Ref.~\onlinecite{PhysRevX.11.011058}.}
\end{figure}
\begin{figure}[b!]
\includegraphics[width=1\linewidth]{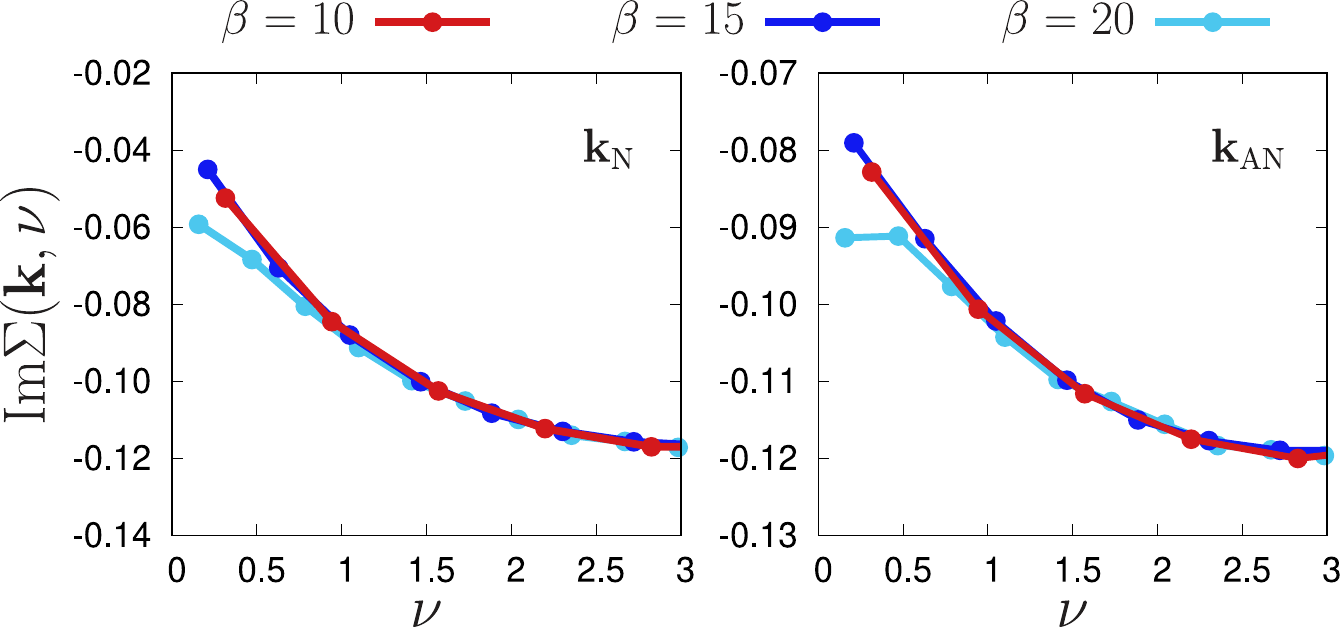}
\caption{\label{fig:pseudogap} Imaginary part of the $\text{D-TRILEX}$ self-energy as the function of Matsubara frequency $\nu$ obtained for $U=2$ at the nodal ${\text{N}=(\pi/2, \pi/2)}$ (left panel) and antinodal ${\text{AN}=(0, \pi)}$ (right panel) points for different inverse temperatures $\beta=10$ (red line), $\beta=15$ (dark blue line), and $\beta=20$ (light blue line).}
\end{figure}

We would like to note that the largest discrepancy between the $\text{D-TRILEX}$ and reference $\text{DiagMC@DF}$ result $\delta = 18\%$ corresponds to the most correlated regime ($U=8$). 
At small and moderate interactions the normalized difference does not exceed 3.5\% ($U=4$). 
At the same time, the maximal difference from the parental LDF is only around 10\% ($U=8$), which can be considered as relatively good result for such a simple theory. 
Finally, we looked at the contribution of the longitudinal particle-particle fluctuations to the $\text{D-TRILEX}$ self-energy~\eqref{eq:Sigma_dual} and, as expected, found it to be negligibly small. 
Indeed, the renormalized singlet interaction in the $\text{D-TRILEX}$ form~\eqref{eq:Ws} does not contain the bare constant interaction $U^{\rm s}$ and therefore describes only the screening of the Coulomb interaction by particle-particle fluctuations.
As the result, we observe that the part of the self-energy that stems from the singlet bosonic mode makes only $3\%$ of the $\text{D-TRILEX}$ self-energy at $U=2$, and does not exceed $1\%$ for other interaction strength. 
Taking into account all above discussions, this result confirms that all particle-particle fluctuations can indeed be safely excluded from the simple $\text{D-TRILEX}$ theory, which however does not hold for every diagrammatic approach.

\begin{figure}[t!]
\includegraphics[width=1.0\linewidth]{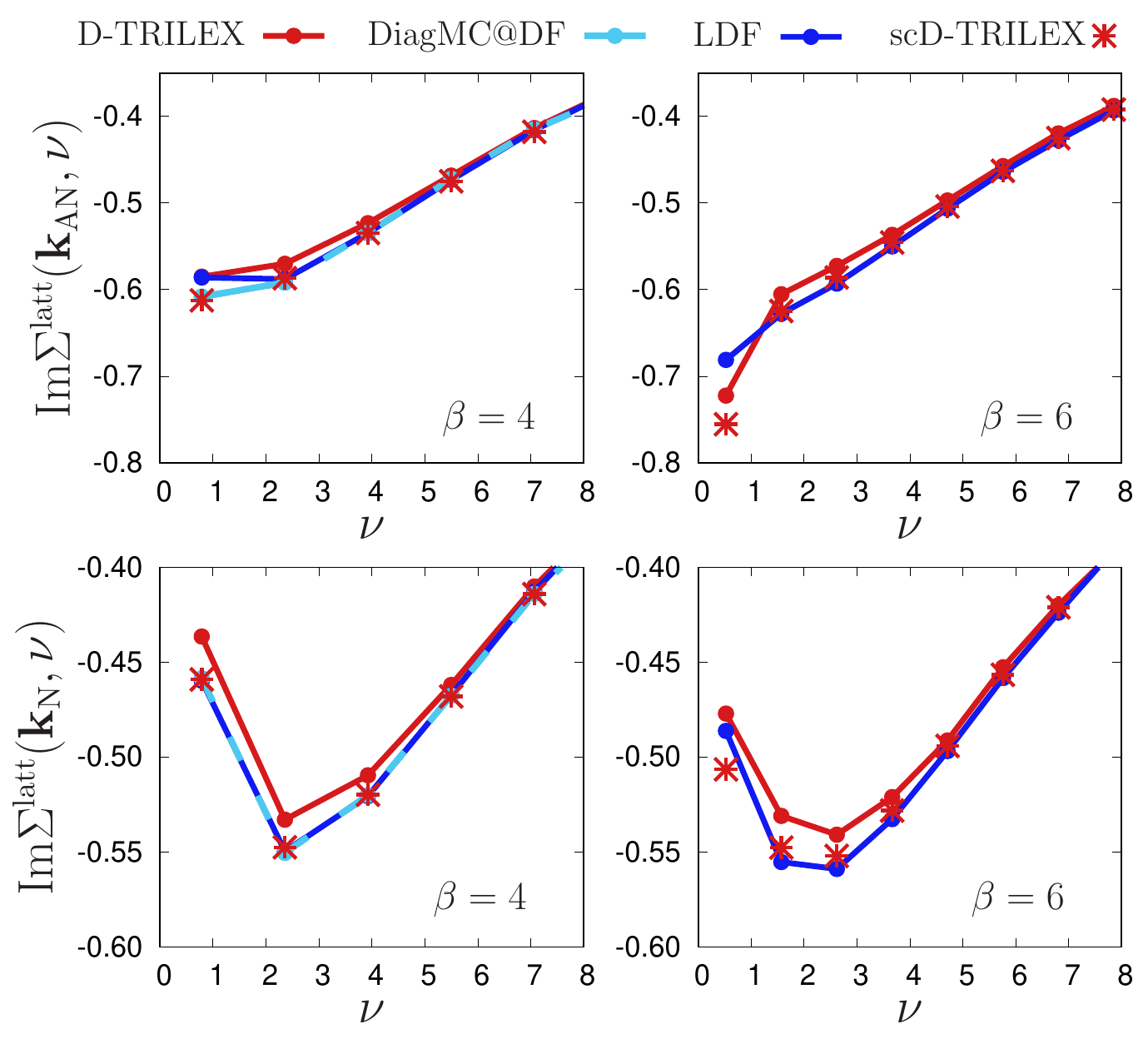} 
\caption{\label{fig:ImSigma_U4} Imaginary part of the lattice self-energy as the function of Matsubara frequency $\nu$ obtained for $U=4$ at the AN (top row) and N (bottom row) points.
Results are calculated at the inverse temperature $\beta=4$ (left column) and $\beta=6$ (right column) using $\text{D-TRILEX}$ (red line), $\text{scD-TRILEX}$ (red stars), and LDF (dark blue line) approaches. The reference $\text{DiagMC@DF}$ result (light blue line) is presented only for $\beta=4$ and is reproduced by the $\text{scD-TRILEX}$ method.}
\end{figure}

\subsection{Low temperature regime}
\label{sec:Low_temp}

In the previous Section we considered only the high-temperature regime ($\beta=2$), where AFM fluctuations are not very strong especially at $U=2$ and $U=4$.
As the next step, we perform calculations at substantially lower temperatures around the DMFT N\'eel point
for the interaction strengths up to a half of the bandwidth ($U\leq4$).
A detailed investigation of the Hubbard model for $U=2$ and different temperatures has been performed in the recent work~\cite{PhysRevX.11.011058}.
This allows for a direct comparison of LDF and $\text{D-TRILEX}$ results with the exact $\text{DiagMC}$ solution of the lattice problem~\eqref{eq:action_latt} presented in that paper.
For $U=4$ we consider only the LDF, the $\text{D-TRILEX}$, and the $\text{DiagMC@DF}$ methods due to the lack of the lattice $\text{DiagMC}$ reference data.
The LDF, the $\text{D-TRILEX}$, and the $\text{DiagMC@DF}$ results are obtained on the basis of the converged DMFT solution in the same way as in the previous Section.
In addition, we also performed fully self-consistent (sc) $\text{D-TRILEX}$ calculations, for which the fermionic hybridization function $\Delta_{\nu}$ of the impurity problem was updated imposing the following self-consistency condition on the local part of the dual Green's function ${\sum_{\bf k}\tilde{G}_{{\bf k}\nu}=0}$ (see e.g. Ref.~\onlinecite{PhysRevB.77.033101}). 

As has been demonstrated in Ref.~\onlinecite{PhysRevX.11.011058}, upon lowering the temperature even a weakly-interacting system goes from a metallic regime, which is characterised by the imaginary part of the self-energy extrapolating to zero at low Matsubara frequencies, 
to a correlated regime where a pseudogap opens.
The latter can be explained by a Slater
mechanism~\cite{PhysRev.82.538} associated with the increase of long-range AFM fluctuations of itinerant electrons. 
The pseudogap opens first at the AN point, which can be detected by the change of the sign in the slope of the self-energy between the first and the second Matsubara frequency.
This N/AN dichotomy appears due to additional suppression of the coherence of single-particle excitations due to the presence of the van Hove singularity at the AN point.

Fig.~\ref{fig:ImSigma_U2} shows that despite a small mismatch at the first Matsubara frequency the LDF and $\text{D-TRILEX}$ methods correctly reproduce the $\text{DiagMC}$ result at $U=2$ and $\beta=10$ above the DMFT N\'eel point ($\beta_{N} \simeq 12.5$).
Moreover, both $\text{D-TRILEX}$ approaches provide identical results for the self-energy, which almost exactly coincides with the one of the LDF approach.
Below the DMFT N\'eel point at $\beta=15$ the $\text{DiagMC}$ self-energy already shows the formation of a pseudogap at the AN point, while the $\text{D-TRILEX}$ and the LDF still exhibit a metallic behaviour.
However, for all other frequencies than the first one the LDF self-energy remains in a very good agreement with the exact result.
In its turn, the $\text{D-TRILEX}$ self-energy starts to deviate from the LDF result, and this discrepancy is more visible at the AN point.
We recall that close to the AFM instability collective fluctuations become strongly anharmonic~\cite{PhysRevB.102.224423}.
In particular, this anharmonicity is significant in the Slater regime of weak interactions, where the magnetic fluctuations are formed by itinerant electrons. 
As a consequence, one can expect that in this regime the transverse modes start to play an important role~\cite{PhysRevB.94.035102, PhysRevB.96.035152, PhysRevB.102.195109}.
We note that in the $\text{D-TRILEX}$ method these transverse contributions are fully discarded, while the LDF approach at least accounts for them in the local four-point vertex function of the impurity problem.
This fact explains why the LDF shows the formation of the pseudogap at the AN point at a bit lower temperature ${T_{\ast}^{\rm AN}=0.059}$ (${\beta=17}$, see Ref.~\onlinecite{harkov2021parameterizations}) than the one of the $\text{DiagMC}$ method ${T_{\ast}^{\rm AN}=0.065}$~\cite{PhysRevX.11.011058}, and the $\text{D-TRILEX}$ approach captures it at ${T_{\ast}^{\rm AN}=0.050}$ (${\beta=20}$, see Fig.~\ref{fig:pseudogap}).
This also explains the result that the most noticeable deviation of the self-energy from the exact result corresponds to the $\text{D-TRILEX}$ method and appears at the AN point where the pseudogap opens first.

Now we increase the strength of the interaction to $U=4$ (Fig.~\ref{fig:ImSigma_U4}) and find that at $\beta=4$ slightly above the DMFT N\'eel point ($\beta_{N}\simeq4.3$) the $\text{D-TRILEX}$ self-energy is in a good agreement with the LDF result for the AN point and shows the beginning of the formation of a pseudogap.
In its turn, the LDF approach agrees with the $\text{DiagMC@DF}$ theory for all frequencies except for the first one.
At the N point the LDF self-energy lies on top of the $\text{DiagMC@DF}$ curve, but the deviation of the $\text{D-TRILEX}$ method from the reference result is more visible.
However, the fully self-consistent calculation strongly improves the $\text{D-TRILEX}$ self-energy, which now perfectly agrees with the $\text{DiagMC@DF}$ result.
At $\beta=6$ below the DMFT N\'eel point the $\text{DiagMC@DF}$ result suffers from the convergence issue~\cite{PhysRevB.94.035102, PhysRevB.96.035152} and is not shown here. 
All other considered methods show an insulating self-energy at the AN point, while the N point remains metallic.
This result confirms the N/AN dichotomy in the formation of a pseudogap.
Also, this result suggests that at moderate interactions collective magnetic fluctuations are less anharmonic in contrast to the weakly-interacting regime.
This fact can be attributed to a more localized behavior of electrons when going away from Slater towards Heisenberg regime of magnetic fluctuations. 
As a consequence, for stronger interactions the formation of the AFM pseudogap can be captured by simpler ladder-like theories.
We also note that at moderate interactions the discrepancy between the $\text{D-TRILEX}$ and the LDF approaches slightly increases upon lowering the temperature. 
However, the self-consistent update of the hybridization function $\Delta_{\nu}$ again improves the agreement between both methods. 
As explicitly stated in Section~\ref{sec:Theory}, in dual theories the hybridization function $\Delta_{\nu}$ is added to the reference system ${\cal S}_{\rm imp}$ and subtracted from the remaining part of the action ${\cal S}_{\rm rem}$ so that the initial problem~\eqref{eq:action_latt} remains unchanged.
Therefore, if the dual problem~\eqref{eq:dual_action} is solved exactly, $\Delta_{\nu}$ can be taken arbitrarily. 
At the same time, any approximate solution depends on the choice for the hybridization function.
In the latter case, the imposed self-consistency condition aims at tuning $\Delta_{\nu}$ in such a way that it accounts for the effect of missing diagrams. 
In this context the fact that at $U=2$ both sc and non-sc $\text{D-TRILEX}$ methods produce identical results demonstrates that the impurity problem of DMFT serves as good reference system in the weakly interacting regime.
On the contrary, already for moderate interaction $U=4$ the self-consistency clearly improves the result of the $\text{D-TRILEX}$ approach, which shows that in this case the DMFT impurity problem does not provide the best possible starting point for partial diagrammatic resummations.

\subsection{Doped regime of the $t-t'$ Hubbard model}

The two-dimensional Hubbard model on a square lattice with nearest-neighbor $t$ and next-nearest-neighbor $t'$ hopping amplitudes is widely known as a prototype model for high-temperature superconducting cuprate compounds. 
The opening of a pseudogap and the dichotomy between the N and AN points in this model has been studied recently in Ref.~\onlinecite{Wu17} in the framework of the exact $\text{DiagMC}$ method. 
There, the authors considered the following set of model parameters $t'=-0.3$, $U=5.6$, $\beta=5$, and $4\%$ hole-doping that leads to a largest onset temperature for the pseudogap.
In our work we address this physically interesting regime for a comparable hole doping of $3.4\%$ within the $\text{scD-TRILEX}$ and $\text{DiagMC@DF}$ approaches.
The obtained self-energies are compared with the exact result of $\text{DiagMC}$ method that was provided by the authors of the Ref.~\onlinecite{Wu17}.
For the sake of consistency, the $\text{DiagMC@DF}$ expansion was performed based on the impurity problem of the $\text{scD-TRILEX}$ approach. 

\begin{figure}[t!]
\centering
\includegraphics[width=1\linewidth]{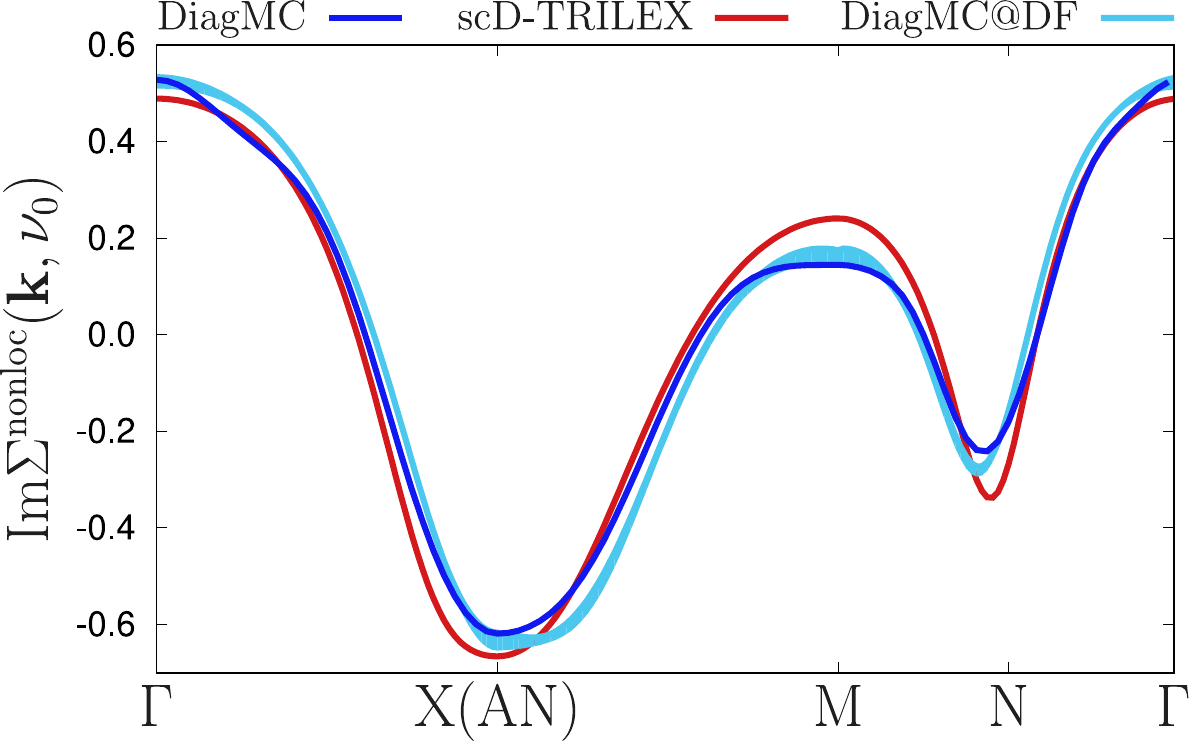}
\caption{\label{fig:U5.6_gmx} Imaginary part of the nonlocal self-energy obtained for the zeroth Matsubara frequency $\nu_0=\pi/\beta$ along the high symmetry path in momentum space ${\bf k}$. Calculations are performed for $U=5.6$, $t'=-0.3$, and $\beta=5$ using $\text{scD-TRILEX}$ (red line) and $\text{DiagMC@DF}$ (light blue line) methods for $3.4\%$ hole-doping. The $\text{DiagMC}$ result (dark blue line) for $4\%$ doping is provided by the authors of Ref.~\onlinecite{Wu17}.}
\end{figure}

In Fig.~\ref{fig:U5.6_gmx} we compare the imaginary part of the nonlocal self-energy $\Sigma^{\rm nonloc}_{{\bf k},\nu_0}$ calculated for the first Matsubara frequency along the high-symmetry path in momentum-space for all three approaches.
To obtain this quantity we subtract the local part $\Sigma^{\rm loc}_{{\bf k},\nu_0}$ from the lattice self energy $\Sigma^{\rm latt}_{{\bf k},\nu_0}$, where ${\Sigma^{\rm loc}_{{\bf k},\nu_0} = \sum_{\bf k}\Sigma^{\rm latt}_{{\bf k},\nu_0}}$.
Due to the lack of reference $\text{DiagMC}$ data, the sum over the Brillouin zone in this expression is approximated by the sum over the high-symmetry path in momentum space.
We find that the nonlocal part of the $\text{DiagMC@DF}$ self-energy is in a very good agreement with the reference $\text{DiagMC}$ result. 
The $\text{scD-TRILEX}$ approach also performs remarkably good in this physically nontrivial regime, especially given that the considered value of the local Coulomb interaction $U=5.6$ exceeds the half of the bandwidth.  
This good agreement in ${\rm Im}\,\Sigma^{\rm nonloc}_{{\bf k},\nu_0}$ indicates that the simple ladder-like $\text{scD-TRILEX}$ method accurately captures the N/AN dichotomy in the formation of a pseudogap in this regime~\cite{Wu17}.
This fact additionally confirms our finding that going away from the Slater regime allows to use less sophisticated methods to capture the effect of collective fluctuations. 

At the same time we find that the $\text{DiagMC@DF}$ and the $\text{scD-TRILEX}$ methods do not provide a good value for the local part of the lattice self-energy.
Indeed, ${\rm Im}\,\Sigma^{\rm loc}_{{\bf k},\nu_0}$ of the $\text{DiagMC@DF}$ calculated for the zeroth Matsubara frequency is equal to $-0.77$.
The corresponding value for the $\text{scD-TRILEX}$ approach is $-0.80$, while the exact $\text{DiagMC}$ result reads $-1.04$.   
This discrepancy can again be explained by the fact that DMFT impurity problem does not provide a good starting point for a diagrammatic expansion already for moderate interactions.
To address this issue, we exploited the dual self-consistency condition to update the fermionic hybridization as an attempt for the improvement of the reference system (see Section~\ref{sec:Low_temp}).
However, the result obtained in this Section clearly demonstrates the need for an even better starting point, which should be able to provide more accurate local quantities to reproduce the exact result.

\section{Conclusion}

To conclude, in this work we investigated the effect of different collective fluctuations on the single-particle properties of correlated electronic systems. 
In order to disentangle local and nonlocal effects, we introduced an effective reference system -- a local impurity problem.
This effective local problem have been solved numerically exactly providing building blocks for a diagrammatic expansion aiming at describing nonlocal correlation effects.
Following the dual fermion/boson idea, we performed this diagrammatic expansion in the dual space truncating the interaction at the two-particle level and thus preserving only the local renormalized four-point vertex function.
Using the partially bosonized representation for this four-point vertex, we investigated the effect of different bosonic modes contributing to the vertex on the electronic self-energy. 
Performing a comprehensive analysis based on DiagMC, $\text{DiagMC@DF}$, LDF, and $\text{D-TRILEX}$ approaches, we have found that irreducible contributions that are not accounted for by the partially bosonized vertex function can be excluded from the theory in a broad range of physical parameters. 
Indeed, they can be completely eliminated in the ladder approximation by a special choice of the bare local interaction in different channels.
In a weakly-interacting regime, the remaining non-ladder contributions have only a minor effect on the electronic self-energy, and at large interactions these contributions are nearly cancelled by transverse singlet fluctuations.
In turn, these transverse singlet modes partially cancel transverse particle-hole fluctuations in weakly- and moderately-interacting regimes.
Finally, longitudinal singlet bosonic modes have been found to be negligibly small in all considered cases.
All these results confirm that in a broad regime of physical parameters the leading contribution to the self-energy is given by the longitudinal particle-hole bosonic modes.
This important statement allows for a drastic simplification of the diagrammatic expansion, which implies a huge reduction of computational efforts.
Consequently, the $\text{D-TRILEX}$ theory, which appears as the result of this simplification, looks as a very promising and powerful tool for solving a broad class of interacting electronic problems.

At the same time, the theory should not be oversimplified. 
Thus, we have shown that considering only second-order dual self-energy does not provide good result even at weak and moderate interactions when the system enters the correlated regime lowering the temperature. 
Instead, the $\text{D-TRILEX}$ method performs remarkably good even below the DMFT N\'eel temperature in the regime where the AFM pseudogap starts to develop.
A good performance of the $\text{D-TRILEX}$ theory has been confirmed in the moderately-correlated half-filled regime of the Hubbard model, and in the case of a $t-t'$ model for hole-doped cuprate compounds.
Importantly, in the latter case we have found that the $\text{D-TRILEX}$ approach provides a reasonably accurate result for the nonlocal part of the self-energy, while the local part is not reproduced correctly. This fact indicates that DMFT does not always provide an optimal way of constructing the local reference (impurity) problem.
We have also found that the simple ladder-like $\text{D-TRILEX}$ theory fails to correctly reproduce the pseudogap formation in the weakly-interacting Slater regime of magnetic fluctuations. 
This can be explained by a strong anharmonicity of collective fluctuations of itinerant electrons close to the AFM instability.
On the contrary, increasing the local Coulomb interaction drives the system away from the Slater regime.
Thus, the electrons become more localized and their collective behavior turns more harmonic, which can be captured by less demanding ladder-like methods.

\begin{acknowledgments}
The authors thank Antoine Georges, Evgeny Kozik, Michel Ferrero, Fedor $\check{\rm S}$imkovic, Riccardo Rossi, and Wei Wu for discussions and for providing the $\text{DiagMC}$ data. The authors also thank Jan Gukelberger for the help with the original $\text{DiagMC@DF}$ code. The work of E.A.S. was supported by the European Union’s Horizon 2020 Research and Innovation programme under the Marie Sk\l{}odowska Curie grant agreement No.~839551 - $\text{2DMAGICS}$. 
The work of A.I.L. is supported by European Research Council via Synergy Grant 854843 - FASTCORR.
M.V., S.B., and A.I.L. acknowledge the support by the Cluster of Excellence ``Advanced Imaging of Matter'' of the Deutsche Forschungsgemeinschaft (DFG) - EXC 2056 - Project No. ID390715994.
The authors also acknowledge the support by North-German Supercomputing Alliance (HLRN) under the Project No. hhp00042.
\end{acknowledgments}

\appendix

\section{Dual boson theory}
\label{app:DB}

The explicit derivation of the DB method can be found in many previous papers on the topic~\cite{Rubtsov20121320, PhysRevB.90.235135, PhysRevB.93.045107, PhysRevB.94.205110, PhysRevB.100.165128, StepanovHarkov}.
However, for the purpose of the current work we have to additionally introduce bosonic variables for the singlet channel.
For this reason here we present a brief derivation of the dual boson theory one more time.
We start with the remaining part of the lattice action~\eqref{eq:action_latt}
\begin{align}
{\cal S}_{\rm rem} = &- \sum_{k,\sigma} c^{*}_{k\sigma} \left[\Delta^{\phantom{*}}_{\nu} - \varepsilon^{\phantom{*}}_{\kv}\right] c^{\phantom{*}}_{k\sigma} 
+ \sum_{q,\vartheta} \xi^{\vartheta} \Bigg\{\rho^{*\,\vartheta}_{q} V^{\vartheta}_{\qv} \, \rho^{\vartheta}_{q} \Bigg\}
\end{align}
Let us perform two Hubbard-Stratonovich transformations
\begin{widetext}
\begin{align}
\exp&\left\{\sum_{k,\sigma} c^{*}_{k\sigma} \left[\Delta^{\phantom{*}}_{\nu} - \varepsilon^{\phantom{*}}_{\kv}\right] c^{\phantom{*}}_{k\sigma}\right\} 
= {\cal D}_{f}\int D[f^{*},f] \exp\left\{-\sum_{k,\sigma} 
\left( f^{*}_{k\sigma} g^{-1}_{\nu\sigma} [\Delta^{\phantom{*}}_{\nu}-\varepsilon^{\phantom{*}}_{\kv}]^{-1} g^{-1}_{\nu\sigma} f^{\phantom{*}}_{k\sigma} + 
f^{*}_{k\sigma} g^{-1}_{\nu\sigma} c^{\phantom{*}}_{k\sigma} 
+ c^{*}_{k\sigma} g^{-1}_{\nu\sigma} f^{\phantom{*}}_{k\sigma}  \right)\right\} \\
\exp&\left\{ - \sum_{q,\vartheta} \xi^{\vartheta} \left(\rho^{*\,\vartheta}_{q} V^{\vartheta}_{\qv} \, \rho^{\vartheta}_{q}\right)\right\} 
= {\cal D}_{\varphi} \int D[\varphi^{\vartheta}] \exp\left\{ \sum_{q,\vartheta} \xi^{\vartheta} \left( \varphi^{*\,\vartheta}_{q} \alpha^{\vartheta\,-1}_{\omega} V^{\vartheta\,-1}_{\qv} \alpha^{\vartheta\,-1}_{\omega} \varphi^{\vartheta}_{q} - 
\varphi^{*\,\vartheta}_{q} \alpha^{\vartheta\,-1}_{\omega} \rho^{\vartheta}_{q}
- \rho^{*\,\vartheta}_{q} \alpha^{\vartheta\,-1}_{\omega} \varphi^{\vartheta}_{q}
\right)\right\}
\end{align}
Here, quantities
$g_{\nu}$ and $w^{\vartheta}_{\omega}$ are the full Green's function and the renormalized interaction of the local impurity problem, respectively, and $\alpha^{\vartheta}_{\omega} = w^{\vartheta}_{\omega}/U^{\vartheta}$.
Terms ${\cal D}_{f} = {\rm det}\left[g_{\nu}\left(\Delta_{\nu}-\varepsilon_{\kv}\right)g_{\nu}\right]$ and ${\cal D}^{-1}_{\varphi} = -\sqrt{{\rm det}\left[\alpha^{\vartheta}_{\omega}V^{\vartheta}_{\qv}\alpha^{\vartheta}_{\omega}\right]}$ can be neglected when calculating expectation values. 
After these transformations the action takes the following form
\begin{align} 
{\cal S}'_{\rm rem}
&= \sum_{i}{\cal S}^{(i)}_{\rm imp} 
+ \sum_{k,\sigma} \left( 
f^{*}_{k\sigma} g^{-1}_{\nu\sigma} c^{\phantom{*}}_{k\sigma}
+ c^{*}_{k\sigma} g^{-1}_{\nu\sigma} f^{\phantom{*}}_{k\sigma} \right) 
+ \sum_{q,\vartheta} \xi^{\vartheta} \left(
\varphi^{*\,\vartheta}_{q} \alpha^{\vartheta\,-1}_{\omega} \rho^{\vartheta}_{q}
+ \rho^{*\,\vartheta}_{q} \alpha^{\vartheta\,-1}_{\omega} \varphi^{\vartheta}_{q} \right) \notag\\
&- \sum_{k,\sigma} f^{*}_{k\sigma} g^{-1}_{\nu\sigma} [\varepsilon^{\phantom{*}}_{\kv}-\Delta^{\phantom{*}}_{\nu}]^{-1} g^{-1}_{\nu\sigma} f^{\phantom{*}}_{k\sigma}
- \sum_{q,\varsigma} \xi^{\vartheta} \left(\varphi^{*\,\varsigma}_{q} \left[\alpha^{\varsigma}_{\omega} V^{\varsigma}_{\qv} \alpha^{\varsigma}_{\omega}\right]^{-1} \varphi^{\varsigma}_{q} \right)
\end{align}
Then, the impurity problem can be integrated out as
\begin{align}
\int D[c^{*},c]\,&\exp\left\{-
\sum_{i}{\cal S}^{(i)}_{\rm imp} 
- \sum_{k,\sigma} \left( 
f^{*}_{k\sigma} g^{-1}_{\nu\sigma} c^{\phantom{*}}_{k\sigma}
+ c^{*}_{k\sigma} g^{-1}_{\nu\sigma} f^{\phantom{*}}_{k\sigma} \right) 
- \sum_{q,\vartheta} \xi^{\vartheta} \left(
\varphi^{*\,\vartheta}_{q} \alpha^{\vartheta\,-1}_{\omega} \rho^{\vartheta}_{q}
+ \rho^{*\,\vartheta}_{q} \alpha^{\vartheta\,-1}_{\omega} \varphi^{\vartheta}_{q} \right)\right\} =
\notag\\
{\cal Z}_{\rm imp} \times &\exp\left\{ - \sum_{k,\sigma} f^{*}_{k\sigma} g^{-1}_{\nu\sigma} f^{\phantom{*}}_{k\sigma} 
- \sum_{q, \vartheta} \xi^{\vartheta} \left(\varphi^{*\,\vartheta}_{q} \alpha^{\vartheta\,-1}_{\omega} \chi^{\vartheta}_{\omega} \alpha^{\vartheta\,-1}_{\omega} \varphi^{\vartheta}_{q}\right)
- \tilde{\cal F}[f,\varphi]
\right\}
\end{align}
\end{widetext}
where ${\cal Z}_{\rm imp}$ and $\chi^{\vartheta}_{\omega}=-\av{\rho^{\vartheta}_{\omega} \, \rho^{\vartheta\,*}_{\omega}}$ are the partition function and the susceptibility of the impurity problem, respectively.
This results in the dual boson action
\begin{align}
{\cal \tilde{S}}
= &-\sum_{k,\sigma} f^{*}_{k\sigma}\tilde{\cal G}^{-1}_{k\sigma}f^{\phantom{*}}_{k\sigma}
- \sum_{q,\vartheta} \xi^{\vartheta} \Bigg\{\varphi^{*\,\vartheta}_{q}
\tilde{\cal W}^{\vartheta\,-1}_{q}
\varphi^{\vartheta}_{q} \Bigg\}
+ \tilde{\cal F}[f,\varphi]
\label{eq:dual_action_app}
\end{align}
The explicit form of the bare fermionic and bosonic propagators of the dual problem is following
\begin{align}
\tilde{\cal G}_{k\sigma} &= g_{\nu\sigma} \left[ [\varepsilon^{\phantom{*}}_{\kv}-\Delta^{\phantom{*}}_{\nu}]^{-1} - g_{\nu\sigma} \right]^{-1} g_{\nu\sigma} = \check{G}_{k\sigma} - g_{\nu\sigma} \\
\tilde{\cal W}^{\vartheta}_{q} &= \alpha^{\vartheta}_{\omega} \left[V^{\vartheta\,-1}_{\qv} - \chi^{\vartheta}_{\omega} \right]^{-1} \alpha^{\vartheta}_{\omega} 
= \check{W}^{\vartheta}_{q} - w^{\vartheta}_{\omega} \label{eq:W_dual_app}
\end{align}
where $\check{G}_{k\sigma}$ and $\check{W}^{\vartheta}_{q}$ are the Green's function and renormalized interaction of EDMFT.

The interaction part of the action $\tilde{\cal F}[f,\varphi]$ being truncated to the two-particle level explicitly reads
\begin{align}
\tilde{\cal F}[f,\varphi]
&\simeq \sum_{q,k,\vartheta}
\xi^{\vartheta} \Bigg\{ \Lambda^{\hspace{-0.05cm}\vartheta}_{\nu\omega} \eta^{*\,\vartheta}_{q,k}
\varphi^{\vartheta}_{q} + \Lambda^{\hspace{-0.05cm}*\,\vartheta}_{\nu\omega} \varphi^{*\,\vartheta}_{q} \eta^{\vartheta}_{q,k} \Bigg\} \notag\\
&+ \frac14 \sum_{q,\{k\},\{\sigma\}} \Gamma_{ph,\,\nu\nu'\omega}^{\sigma\sigma'\sigma''\sigma'''} f^{*}_{k\sigma} f^{\phantom{*}}_{k+q, \sigma'} f^{*}_{k'+q, \sigma'''} f^{\phantom{*}}_{k'\sigma''}
\label{eq:lowestint}
\end{align}
where, $\eta^{\theta}_{q,k}$ have been defined in Eqs.~\eqref{eq:eta_dm},~\eqref{eq:eta_s}, and~\eqref{eq:eta_sast}.
The four-point vertex functions in the particle-hole $\Gamma_{ph}$ and particle-particle $\Gamma_{pp}$ form are defined as
\begin{align}
\Gamma_{ph,\,\nu\nu'\omega}^{\,\sigma_1\sigma_2\sigma_3\sigma_4} 
&= \frac{\av{c^{\phantom{*}}_{\nu\sigma_1} c^{*}_{\nu+\omega,\sigma_2} c^{*}_{\nu',\sigma_3} c^{\phantom{*}}_{\nu'+\omega,\sigma_4}}_{\rm c}}{
g_{\nu\sigma_1} g_{\nu+\omega,\sigma_2} g_{\nu'\sigma_3} g_{\nu'+\omega,\sigma_4}} \notag\\
\Gamma_{pp,\,\nu\nu'\omega}^{\,\sigma_1\sigma_2\sigma_3\sigma_4} 
&= \frac{\av{c^{\phantom{*}}_{\nu\sigma_1} c^{\phantom{*}}_{\omega-\nu,\sigma_2} c^{*}_{\nu'\sigma_3} c^{*}_{\omega-\nu',\sigma_4}}_{\rm c}}{ g_{\nu\sigma_1} g_{\omega-\nu,\sigma_2} g_{\nu'\sigma_3} g_{\omega-\nu',\sigma_4}} \label{eq:Gamma_appendix}
\end{align}
where $\av{\ldots}_{\rm c}$ denotes the connected part of the correlation function.
The following relation between two representations holds 
\begin{align}
\Gamma_{pp,\,\nu\nu'\omega}^{\,\sigma_1\sigma_2\sigma_3\sigma_4} = \Gamma_{ph,\,\nu,\omega-\nu',\nu'-\nu}^{\,\sigma_1\sigma_3\sigma_4\sigma_2} = -\,\Gamma_{ph,\,\nu,\nu',\omega-\nu-\nu'}^{\,\sigma_1\sigma_4\sigma_3\sigma_2}
\end{align}
Density (d), magnetic (m), singlet (s), and triplet (t) components of the four-point vertex are defined as
\begin{align}
\Gamma^{\,\rm d/m}_{\nu\nu'\omega} &= \Gamma_{ph,\,\nu\nu'\omega}^{\,\uparrow\uparrow\uparrow\uparrow} \pm \Gamma_{ph,\,\nu\nu'\omega}^{\,\uparrow\uparrow\downarrow\downarrow} \notag\\
\Gamma^{\,\rm s/t}_{\nu\nu'\omega} &= \frac12 \Gamma_{pp,\,\nu\nu'\omega}^{\,\uparrow\downarrow\uparrow\downarrow} \mp 
\frac12 \Gamma_{pp,\,\nu\nu'\omega}^{\,\uparrow\downarrow\downarrow\uparrow}
\end{align}
The three-point interactions in the channel representation are defined as 
\begin{align}
\Lambda_{\nu\omega}^{\hspace{-0.05cm}\varsigma} &=  
\frac{\av{c^{\phantom{*}}_{\nu\uparrow} c^{*}_{\nu+\omega\uparrow}\,\rho^{*\,\varsigma}_{\omega}}}{g_{\nu\uparrow} g_{\nu+\omega\uparrow}\alpha^{\varsigma}_{\omega}}; ~~~~
\Lambda_{\nu\omega}^{\hspace{-0.05cm}*\,\varsigma} =  
\frac{\av{c^{\phantom{*}}_{\nu+\omega\uparrow} c^{*}_{\nu\uparrow}\,\rho^{\varsigma}_{\omega}}}{g_{\nu+\omega\uparrow} g_{\nu\uparrow}\alpha^{\varsigma}_{\omega}}; \notag\\
\Lambda_{\nu\omega}^{\hspace{-0.05cm}\text{s}} &= \frac{
\av{c^{\phantom{*}}_{\nu\uparrow} c^{\phantom{*}}_{\omega-\nu\downarrow}\,\rho^{*\,\text{s}}_{\omega}}}{g_{\nu\uparrow} g_{\omega-\nu\downarrow}\alpha^{\text{s}}_{\omega}}; ~~~~~
\Lambda_{\nu\omega}^{\hspace{-0.05cm}*\,\text{s}} = \frac{
\av{c^{*}_{\omega-\nu\downarrow} c^{{*}}_{\nu\uparrow}\,\rho^{\text{s}}_{\omega}}}{g_{\omega-\nu\downarrow} g_{\nu\uparrow} \alpha^{\text{s}}_{\omega}}.
\label{eq:Lambda_app}
\end{align}
In the particle-hole channel the three-point vertex obeys the useful relation
$\Lambda^{\hspace{-0.05cm}*\,\varsigma}_{\nu\omega} = \Lambda^{\hspace{-0.05cm}\varsigma}_{\nu+\omega,-\omega}$.
The three-point vertex in the triplet channel is not introduced, because the composite variable $\rho^{\rm t}$ is identically zero in the single-band case.

It is important to note that in dual diagrams the bosonic line always connects two three-point vertex functions. 
Using Eqs.~\eqref{eq:W_dual_app} and~\eqref{eq:Lambda_app}, one finds that 
\begin{align}
\Lambda^{\hspace{-0.05cm}\vartheta}_{\nu\omega} \tilde{W}^{\vartheta}_{q} \Lambda^{\hspace{-0.05cm}*\,\vartheta}_{\nu\omega} \sim 
\alpha^{\vartheta\,-1}_{\nu\omega} \tilde{W}^{\vartheta}_{q} \alpha^{\vartheta\,-1}_{\nu\omega} = \left[V^{\vartheta\,-1}_{\qv} - \chi^{\vartheta}_{\omega} \right]^{-1}
\label{eq:combination}
\end{align}
This relation shows that $\alpha^{\vartheta}_{\omega}$, which is the only quantity that explicitly contains the bare local interaction $U^{\vartheta}$, drops out from the dual diagrammatics.
Therefore, physical observables that can be found via the exact relation between correlation function written in terms of dual $f$ and original $c$ fermion variables also do not depend on the choice of the bare interaction $U^{\theta}$ for different bosonic channels.
This fact is not surprising, because the on-site Coulomb interaction $U$ is fully accounted by the impurity problem and is already contained in the four-point vertex.\\

\section{D-TRILEX theory}
\label{app:D-TRILEX}

In this section we explicitly show the transformation that reduces the dual boson problem~\eqref{eq:dual_action} to the fermion-boson action~\eqref{eq:fbaction} of the $\text{D-TRILEX}$ theory. 
The key idea is to find such Hubbard-Stratonovich transformation that produces the interaction in the partially bosonized form of Eq.~\eqref{eq:Gamma} that nearly cancels the exact four-point vertex function from the theory. 
To this aim let us first add and subtract the term 
\begin{align}
{\sum_{q, \vartheta} \xi^{\vartheta} \left(\varphi^{*\,\vartheta}_{q} \bar{w}_{\omega}^{\vartheta\,-1} \varphi^{\vartheta}_{q}\right)} 
\label{eq:w_arb_app}
\end{align}
from the dual action~\eqref{eq:dual_action_app}. 
At this step one can consider $\bar{w}^{\vartheta}_{\omega}$ as an arbitrary function that will be determined later.
This procedure will allows us to integrate out dual bosonic fields $\varphi$ with respect to the arbitrary Gaussian part of the action introduced in Eq.~\eqref{eq:w_arb_app}. 
To illustrate this point, we make use of following Hubbard-Stratonovich transformations
\begin{widetext}
\begin{align}
&\exp\left\{
\sum_{q,\vartheta} \xi^{\vartheta} \varphi^{*\,\vartheta}_{q}
\left[\tilde{W}^{\vartheta\,-1}_{q} + \bar{w}^{\vartheta\,-1}_{\omega} \right]
\varphi^{\vartheta}_{q}
\right\} = \notag\\
&{\cal D}_{b}
\int D[b^{\vartheta}] \exp\left\{-\sum_{q,\vartheta} \xi^{\vartheta} \Bigg( b^{*\,\vartheta}_{q} \bar{w}^{\vartheta\,-1}_{\omega}
\left[\tilde{W}^{\vartheta\,-1}_{q} + \bar{w}^{\vartheta\,-1}_{\omega} \right]^{-1} \bar{w}^{\vartheta\,-1}_{\omega} b^{\vartheta}_{q} 
- \varphi^{*\,\vartheta}_{q} \bar{w}^{\vartheta\,-1}_{\omega} b^{\vartheta}_{q}
- b^{*\,\vartheta}_{q} \bar{w}^{\vartheta\,-1}_{\omega} \varphi^{\vartheta}_{q} \Bigg)\right\}
\end{align}
where ${\cal D}^{-1}_{b} = \sqrt{{\rm det}\left[\bar{w}^{\vartheta}_{\omega}
\left(\tilde{W}^{\vartheta\,-1}_{q} + \bar{w}^{\vartheta\,-1}_{\omega} \right) \bar{w}^{\vartheta}_{\omega}\right]}$. 
The action transforms to
\begin{align}
{\cal \tilde{S}}' =
&-\sum_{k,\sigma} f^{*}_{k\sigma}\tilde{\cal G}^{-1}_{k\sigma}f^{\phantom{*}}_{k\sigma}
+ \sum_{q,\vartheta} \xi^{\vartheta} \Bigg\{ b^{*\,\vartheta}_{q} \bar{w}^{\vartheta\,-1}_{\omega}
\left[\tilde{W}^{\vartheta\,-1}_{q} + \bar{w}^{\vartheta\,-1}_{\omega} \right]^{-1} \bar{w}^{\vartheta\,-1}_{\omega} b^{\vartheta}_{q} \Bigg\} 
+ \sum_{q, \vartheta} \xi^{\vartheta} \Bigg\{ \varphi^{*\,\vartheta}_{q} \bar{w}_{\omega}^{\vartheta\,-1} \varphi^{\vartheta}_{q} 
- \varphi^{*\,\vartheta}_{q} \bar{w}^{\vartheta\,-1}_{\omega} b^{\vartheta}_{q} - b^{*\,\vartheta}_{q} \bar{w}^{\vartheta\,-1}_{\omega} \varphi^{\vartheta}_{q} \Bigg\}
+ \tilde{\cal F}[f,\varphi]
\end{align}
Finally, dual bosonic fields $\varphi$ can be integrated out with respect to the new Gaussian part of the dual action as
\begin{align}
\int & D[\varphi^{\varsigma}]\,\exp\left\{-\sum_{q,k,\vartheta} \xi^{\vartheta} \Bigg( \varphi^{*\,\vartheta}_{q} \bar{w}_{\omega}^{\vartheta\,-1} \varphi^{\vartheta}_{q} 
- \varphi^{*\,\vartheta}_{q} \left[ \bar{w}^{\vartheta\,-1}_{\omega} b^{\vartheta}_{q} - \Lambda^{\hspace{-0.05cm}*\,\vartheta}_{\nu\omega} \eta^{\vartheta}_{q,k} \right]
- \left[b^{*\,\vartheta}_{q} \bar{w}^{\vartheta\,-1}_{\omega} - \eta^{*\,\vartheta}_{q,k} \Lambda^{\hspace{-0.05cm}\vartheta}_{\nu\omega} \right]
\varphi^{\vartheta}_{q} \Bigg) \right\} = \notag\\
& {\cal Z}_{\varphi} \times \exp\left\{ \sum_{q,k,\vartheta}
\xi^{\vartheta} \Bigg(b^{*\,\vartheta}_{q} \bar{w}_{\omega}^{\vartheta\,-1} b^{\vartheta}_{q} 
- \Lambda^{\hspace{-0.05cm}\vartheta}_{\nu\omega} \eta^{*\,\vartheta}_{q,k}
b^{\vartheta}_{q} - \Lambda^{\hspace{-0.05cm}*\,\vartheta}_{\nu\omega} b^{*\,\vartheta}_{q} \eta^{\vartheta}_{q,k} 
+ \eta^{*\,\vartheta}_{q,k}\Lambda^{\hspace{-0.05cm}\vartheta}_{\nu\omega} \bar{w}^{\vartheta}_{\omega} \Lambda^{\hspace{-0.05cm}*\,\vartheta}_{\nu\omega} \eta^{\vartheta}_{q,k} \Bigg) \right\}
\label{eq:HS_dms_app}
\end{align}
\end{widetext}
where ${\cal Z}_{\varphi}$ is a partition function of the Gaussian part of the bosonic action. 
Being written in the antisymmetrized form, the quartic term $\eta^{*\,\vartheta}_{q,k}\Lambda^{\hspace{-0.05cm}\vartheta}_{\nu\omega} \bar{w}^{\vartheta}_{\omega} \Lambda^{\hspace{-0.05cm}*\,\vartheta}_{\nu\omega} \eta^{\vartheta}_{q,k}$ in Eq.~\eqref{eq:HS_dms_app} makes the partially bosonized representation for the four-point vertex specified in Eq.~\eqref{eq:Gamma}~\cite{StepanovHarkov}.
Since this effective vertex function is generated with the opposite sign, it cancels the exact four-point vertex if $\bar{w}^{\vartheta}_{\omega}$ is defined as in Eqs.~\eqref{eq:w_dm} and~\eqref{eq:w_s}. 
After that, the dual problem reduces to the action of the $\text{D-TRILEX}$ theory~\eqref{eq:fbaction}.

\twocolumngrid

\section{Relation between DB and D-TRILEX diagrams}
\label{app:Sigma}

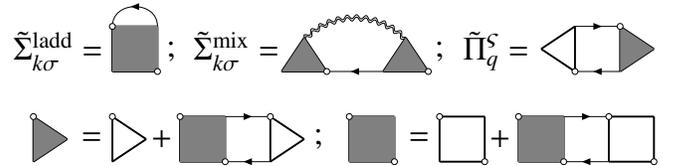
\begin{figure}[b!]
\scalebox{.6}{
\begin{tikzpicture}
\begin{feynman}[small]
    \vertex (b1);
    \vertex[right=2cm of b1] (g1);

    \vertex[left=1.2cm of b1] (b-1) ;
    \vertex[right=2.5cm of b1] (b-2) ;
    \vertex[below=0.55cm of b-1] (b00);
    \vertex[below=0.0001cm of b-1] (b0)
    {\huge{$\tilde{\Sigma}^{\rm ladd}_{k\sigma}=$}};
    \vertex[right=2.5cm of b0] (b-3)
    {\huge{;}};
    
    \vertex[right=1cm of b1] (g2);
    \vertex[below=1cm of g2] (g3);
    \vertex[below=1cm of b1] (g4);
    
    \diagram* [very thick]{
      {
        (b1) -- (g2) -- (g3) -- (g4) -- (b1),
      },
      };
    \fill [black!50] (b1) -- (g2) -- (g3) -- (g4) -- cycle;
      \diagram* {
      {[edges = fermion]
        (g2) -- [half right] (b1),
      },
      };
    
    \draw[fill=white] (b1) circle(0.65mm);
    \draw[fill=white] (g3) circle(0.65mm);


    \vertex[right=2.75cm of b1]  (a1);
    \vertex[below=0.0cm of a1]  (c2)
    {\huge{$\tilde{\Sigma}^{\rm mix}_{k\sigma}=$}};
    \vertex[right=4.5cm of c2] (b-3)
    {\huge{;}};
    \vertex[right=1.0cm of a1] (h1);
    \vertex[below=0.25cm of h1] (h1);
    \vertex[right=0.5cm of h1] (h2);
    \vertex[below=0.75cm of h1] (h3);
    \vertex[right=1cm of h3] (h4);
    
    \vertex[right=1.25cm of h4] (h5);
    \vertex[above=0.75cm of h5] (h6);
    \vertex[right=0.5cm of h6] (h7);
    \vertex[right=1cm of h5] (h8);
    
    \vertex[below=0.04cm of h2] (h11);
    \vertex[right=0.03cm of h11] (h9);
    \vertex[below=0.04cm of h7] (h11);
    \vertex[left=0.03cm of h11] (h10);
    
    \diagram* [very thick]{
      {
        (h2) -- (h3) -- (h4) -- (h2),
      },
      {
       (h5) -- (h7) -- (h8) -- (h5)
      },
      };
    \diagram* {
      {[edges = fermion]
        (h5) -- (h4),
      },
      };
    \diagram* {
      {[edges=boson]
        (h7) -- [quarter right] (h2),
      },
      {[edges=boson]
        (h10) -- [quarter right] (h9),
      },
      };
    \fill [black!50] (h2) -- (h3) -- (h4) -- cycle;  
    \fill [black!50] (h5) -- (h7) -- (h8) -- cycle;  
    \draw[fill=white] (h4) circle(0.65mm);
    \draw[fill=white] (h8) circle(0.65mm);

    \vertex[right=9.5cm of b1]  (b1);

    \vertex[left=1.0cm of b1] (b-1) ;
    \vertex[right=2.5cm of b1] (b-2) ;
    \vertex[below=0.55cm of b-1] (b00);
    \vertex[below=0.0001cm of b-1] (b0)
    {\huge{$\tilde{\Pi}^{\varsigma}_{q}=$}};

    \vertex[right=0.75cm of b1] (b2);
    \vertex[below=1cm of b2] (b3);
    \vertex[below=0.5cm of b1] (b4);
    
    \vertex[right=1.0cm of b2] (b5);
    \vertex[right=0.75cm of b5] (b6);
    \vertex[below=0.5cm of b6] (b7);
    \vertex[below=1cm of b5] (b8);

    \diagram* [very thick]{
      {
       (b2) -- (b3) -- (b4) -- (b2),
      },
      {
        (b5) -- (b7) -- (b8) -- (b5),
      },
      };
      \diagram* {
      {[edges = fermion]
        (b2) -- (b5),
      },
      {[edges = fermion]
        (b8) -- (b3),
      },
      };
    
    \fill [black!50] (b5) -- (b7) -- (b8) -- cycle;
    \draw[fill=white] (b3) circle(0.65mm);
    \draw[fill=white] (b5) circle(0.65mm);
    
    \vertex[below=2.0cm of b1] (b1);
    \vertex[left=9.5cm of b1] (b1);
    \vertex[right=1.5cm of b1] (g1);
    \vertex[left=1.75cm of b1] (a1);
    \vertex[right=0.75cm of a1] (a2);
    \vertex[below=0.5cm of a2] (a3);
    \vertex[below=1cm of a1] (a4); 

    \vertex[left=1.4cm of b1] (b-1) ;
    \vertex[right=2.5cm of b1] (b-2) ;
    \vertex[below=0.5cm of b-1] (b0);
    \vertex[right=0.6cm of b0] (b00)
    {\huge{$=$}};
    
    \vertex[right=0.75cm of b1] (g2);
    \vertex[below=0.5cm of g2] (g3);
    \vertex[below=1cm of b1] (g4);
    \vertex[right=0.0cm of g3] (g2)
    {\huge{$+$}};
    \vertex[right=3.6cm of g3] (b-3)
    {\huge{;}};
    
    \vertex[right=1cm of g1] (b2);
    \vertex[below=1cm of b2] (b3);
    \vertex[below=1cm of g1] (b4);
    
    \vertex[right=1.0cm of b2] (b5);
    \vertex[right=0.75cm of b5] (b6);
    \vertex[below=0.5cm of b6] (b7);
    \vertex[below=1cm of b5] (b8); 
    
    \diagram* [very thick]{
      {
        (a1) -- (a3) -- (a4) --(a1),
      },

      };
    
    \diagram* [very thick]{
      {
        (b1) -- (g3) -- (g4) -- (b1),
      },

      };
    
    \diagram* [very thick]{
      {
        (g1) -- (b2) -- (b3) -- (b4) -- (g1),
      },
      {
        (b5) -- (b7) -- (b8) -- (b5),
      },
      };
      
    \fill [black!50] (a1) -- (a3) -- (a4) -- cycle;
    \fill [black!50] (g1) -- (b2) -- (b3) -- (b4) -- cycle;
      \diagram* {
      {[edges = fermion]
        (b2) -- (b5),
      },
      {[edges = fermion]
        (b8) -- (b3),
      },
      };
    
    \draw[fill=white] (b1) circle(0.65mm);
    \draw[fill=white] (a1) circle(0.65mm);  
    \draw[fill=white] (g1) circle(0.65mm);
    \draw[fill=white] (b3) circle(0.65mm);
    \draw[fill=white] (b5) circle(0.65mm);

    \vertex[right=7.25cm of b1] (b1);
    \vertex[right=1.75cm of b1] (g1);
    \vertex[left=2.0cm of b1] (a1);
    \vertex[right=1.0cm of a1] (a2);
    \vertex[below=1cm of a2] (a3);
    \vertex[below=1cm of a1] (a4); 

    \vertex[left=1.8cm of b1] (b-1) ;
    \vertex[right=2.5cm of b1] (b-2) ;
    \vertex[below=0.5cm of b-1] (b0);
    \vertex[right=1.0cm of b0] (b00)
    {\huge{$=$}};
    
    \vertex[right=1cm of b1] (g2);
    \vertex[below=1cm of g2] (g3);
    \vertex[below=1cm of b1] (g4);
    \vertex[below=0.5cm of b1] (g5);
    \vertex[right=1cm of g5] (g6)
    {\huge{$+$}};
    
    \vertex[right=1cm of g1] (b2);
    \vertex[below=1cm of b2] (b3);
    \vertex[below=1cm of g1] (b4);
    
    \vertex[right=1.0cm of b2] (b5);
    \vertex[right=1.0cm of b5] (b6);
    \vertex[below=1.0cm of b6] (b7);
    \vertex[below=1cm of b5] (b8); 
    
    \diagram* [very thick]{
      {
        (a1) -- (a2) -- (a3) -- (a4) --(a1),
      },

      };
    
    \diagram* [very thick]{
      {
        (b1) -- (g2) -- (g3) -- (g4) -- (b1),
      },

      };
    
    \diagram* [very thick]{
      {
        (g1) -- (b2) -- (b3) -- (b4) -- (g1),
      },
      {
        (b5) -- (b6) -- (b7) -- (b8) -- (b5),
      },
      };
      
    \fill [black!50] (a1) -- (a2) -- (a3) -- (a4) -- cycle;
    \fill [black!50] (g1) -- (b2) -- (b3) -- (b4) -- cycle;
      \diagram* {
      {[edges = fermion]
        (b2) -- (b5),
      },
      {[edges = fermion]
        (b8) -- (b3),
      },
      };

    \draw[fill=white] (a1) circle(0.65mm);
    \draw[fill=white] (a3) circle(0.65mm);
    \draw[fill=white] (b1) circle(0.65mm);
    \draw[fill=white] (g3) circle(0.65mm);
    \draw[fill=white] (g1) circle(0.65mm);
    \draw[fill=white] (b3) circle(0.65mm);
    \draw[fill=white] (b5) circle(0.65mm);
    \draw[fill=white] (b7) circle(0.65mm);
    
    \end{feynman}
  \end{tikzpicture}
  }
  \caption{\label{fig:Sigma_Pi_app} Top row: ladder $\tilde{\Sigma}^{\rm ladd}_{k\sigma}$~\eqref{eq:Sigma_ladd} and mixed $\tilde{\Sigma}^{\rm mix}_{k\sigma}$~\eqref{eq:Sigma_mix} contributions to the LDB self-energy, and the LDB polarization operator $\tilde{\Pi}^{\varsigma}_{q}$~\eqref{eq:DualPi}. Bottom row: Screened three-point $L^{\hspace{-0.05cm}\vartheta}_{\nu q}$~\eqref{eq:Lvertex} and four-point ${\rm P}^{\,\vartheta}_{\nu\nu'q}$~\eqref{eq:Pvertex} vertex functions in the LDB approximation. }
\end{figure}

In this section we establish the relation between ladder DB and $\text{D-TRILEX}$ diagrams for the self-energy. As shown in Ref.~\onlinecite{PhysRevB.93.045107}, the LDB diagrams for the self-energy and polarization operator can be obtained form the dual functional, which yields  
\begin{align}
\tilde{\Sigma}^{\rm LDB}_{k\sigma} =
\tilde{\Sigma}^{\rm LDF}_{k\sigma} +
\tilde{\Sigma}^{\rm mix}_{k\sigma}
\label{eq:Sigma_LDB}
\end{align}
The ladder DF self-energy
\begin{align}
\tilde{\Sigma}^{\rm LDF}_{k\sigma} =
\tilde{\Sigma}^{\rm ladd}_{k\sigma} -\tilde{\Sigma}^{(2)}_{k\sigma}
\label{eq:Sigma_LDF}
\end{align}
is given by the two-particle ladder diagram
\begin{align}
\tilde{\Sigma}^{\rm ladd}_{k\sigma} = -\sum_{q,k',\{\sigma\}}\tilde{G}^{\phantom{f}}_{k+q,\sigma'}{\rm P}^{\,\sigma\sigma'\sigma''\sigma'''}_{ph,\,\nu\nu'q}\delta_{kk'}\delta_{\sigma\sigma''}\delta_{\sigma'\sigma'''}
\label{eq:Sigma_ladd}
\end{align}
As follows from the Schwinger-Dyson for the dual self-energy, the ladder diagram accounts for twice the contribution of the second-order self-energy~\cite{hafermann2010numerical}
\begin{align}
\Sigma_{k\sigma}^{(2)} &= -\frac{1}{2}\sum_{q, k',\{\sigma\}} \Gamma^{\,\sigma \sigma' \sigma'' \sigma'''}_{ph,\,\nu\nu'\omega} \tilde{G}_{k',\sigma''} \tilde{G}_{k'+q, \sigma'''} \tilde{G}_{k+q,\sigma'} \Gamma^{\,\sigma'' \sigma''' \sigma\sigma'}_{ph,\,\nu'\nu\omega}
\label{Sigma_2nd}
\end{align}
which has to be excluded from the expression~\eqref{eq:Sigma_LDF} in order to avoid the double-counting.
The mixed diagram that additionally appears in the DB theory due to the presence of the bosonic propagator $\tilde{W}^{\varsigma}_{q}$ is as follows:
\begin{align}
\tilde{\Sigma}^{\rm mix}_{k\sigma} = -\sum_{q,\varsigma} L^{\hspace{-0.05cm}\varsigma}_{\nu q} \tilde{G}_{q+k,\sigma} \tilde{W}^{\varsigma}_{q} L^{\hspace{-0.05cm}*\,\varsigma}_{\nu q}
\label{eq:Sigma_mix}
\end{align}
Dressed dual fermionic and bosonic propagators can be found using corresponding Dyson equations 
\begin{align}
\tilde{G}_{k\sigma}^{-1} &= \tilde{\cal G}_{k\sigma}^{-1} -  \tilde{\Sigma}_{k\sigma}\\
\tilde{W}_{q}^{\varsigma~-1} &= \tilde{\cal W}_{q}^{\varsigma~-1} - \tilde{\Pi}^{\varsigma}_{q}
\label{eq:DualW_Dyson}
\end{align}
where the dual polarization operator in the ladder approximation reads
\begin{align}
\tilde{\Pi}^{\varsigma}_{q}
= \sum_{k,\sigma} \Lambda^{\hspace{-0.05cm}*\,\varsigma}_{\nu \omega} \tilde{G}_{k\sigma} \tilde{G}_{q+k,\sigma} L^{\hspace{-0.05cm}\varsigma}_{\nu q}
\label{eq:DualPi}
\end{align}
The screened three- and four-point vertices in the horizontal particle-hole (charge and spin) and particle-particle (singlet) channels are
\begin{align}
L^{\hspace{-0.05cm}\vartheta}_{\nu q} &= \Lambda^{\hspace{-0.05cm}\vartheta}_{\nu \omega} + \sum_{k_1} {\rm P}^{\,\vartheta}_{\nu \nu_1 q} \tilde{X}^{0\,\vartheta}_{k_1,q} \Lambda^{\hspace{-0.05cm}\vartheta}_{\nu_1 \omega}
\label{eq:Lvertex}\\
{\rm P}^{\,\vartheta}_{\nu\nu'q} &= 
\Gamma^{\,\vartheta}_{\nu\nu'\omega} + 
\sum_{k_1} {\rm P}^{\,\vartheta}_{\nu \nu_1 q} \tilde{X}^{0\,\vartheta}_{k_1,q}
\Gamma^{\,\vartheta}_{\nu_1 \nu' \omega} 
\label{eq:Pvertex}
\end{align}
Here, $\tilde{X}^{0\,\varsigma}_{k,q} = \tilde{G}^{\phantom{\varsigma}}_{k\sigma} \tilde{G}^{\phantom{\varsigma}}_{k+q, \sigma}$ and $\tilde{X}^{0\,\rm s}_{k,q} = - \, \tilde{G}^{\phantom{\varsigma}}_{k\uparrow} \tilde{G}^{\phantom{\varsigma}}_{q-k, \downarrow}$.
The screened vertices in the vertical $\overline{\rm P}^{\,\varsigma}$ and horizontal ${\rm P}^{\,\varsigma}$ particle-hole channels are connected via the relation ${\overline{\rm P}^{\,\varsigma}_{kk'\omega} = -\,{\rm P}^{\,\varsigma}_{\nu,\nu+\omega,k'-k}}$.
Note that the DB theory does not account for fluctuations in the particle-particle channel, because they are negligibly small in the ladder approximation.
Therefore, the LDB self-energy~\eqref{eq:Sigma_LDB} contains the three- and four-point vertices that are screened only in the particle-hole ($\varsigma$) channel.
Diagrammatic expressions for the LDB self-energy and polarization operator, as well as for the screened three- and four-point vertices, are shown in Fig.~\ref{fig:Sigma_Pi_app}.

Let us derive $\text{D-TRILEX}$ diagrams for the self-energy as an approximation of the ladder DB theory. To this aim one can use the partially bosonized representation for the four-point vertex function~\eqref{eq:Gamma} and keep only longitudinal bosonic fluctuations.

\subsubsection{First-order self-energy}

The first-order contribution to the ladder part of the self-energy~\eqref{eq:Sigma_ladd} written in the channel representation becomes
\begin{align}
\tilde{\Sigma}^{(1)}_{\nu\sigma} &= -\sum_{q,\sigma'}\tilde{G}^{\phantom{f}}_{k+q,\sigma'}\Gamma^{\,\sigma\sigma'\sigma\sigma'}_{\nu\nu\omega} \notag\\
&= -\frac12 \sum_{q,\varsigma} \tilde{G}^{\phantom{f}}_{k+q,\sigma}\Gamma^{\,\varsigma}_{\nu\nu\omega} \notag\\
&\simeq -\sum_{q,\varsigma} \tilde{G}^{\phantom{f}}_{k+q,\sigma} M^{\varsigma}_{\nu\nu\omega} 
+ \sum_{q} \tilde{G}^{\phantom{f}}_{q-k,\sigma} M^{\rm s}_{\nu\nu\omega} \notag\\
&= -\sum_{q,\varsigma} \left\{ \Lambda^{\hspace{-0.05cm}\varsigma}_{\nu\omega} \tilde{G}_{q+k,\sigma} \bar{w}^{\varsigma}_{\omega} \Lambda^{\hspace{-0.05cm}*\,\varsigma}_{\nu\omega} 
- \Lambda^{\hspace{-0.05cm}\rm s}_{\nu\omega} \tilde{G}_{q-k,\overline{\sigma}} \bar{w}^{\rm s}_{\omega} \Lambda^{\hspace{-0.05cm}*\,\rm s}_{\nu\omega} \right\}
\label{eq:Sigma_1st}
\end{align}
Here, a trivial Hartree-like contribution to the dual self-energy $\tilde{\Sigma}^{\rm H}_{\nu\sigma} = 2\Lambda^{\rm d}_{\nu,0}\bar{w}^{\rm d}_{0}\sum_{k'}\Lambda^{\rm d}_{\nu',0}\tilde{G}_{k'\sigma}$ that appears due to $M_{\nu,\nu+\omega,\nu'-\nu}$ terms in Eq.~\eqref{eq:Gamma} have been omitted for simplicity. 

\subsubsection{Second-order self-energy}

The second-order self-energy~\eqref{Sigma_2nd} can be simplified in the same way. Eq.~\eqref{Sigma_2nd} can first be rewritten in the channel representation as
\begin{align}
\tilde{\Sigma}_{k,\sigma_1}^{(2)} &=-\frac{1}{4}\sum_{k', q, \varsigma} \tilde{G}^{\phantom{\varsigma}}_{k+q} \Gamma_{\nu\nu'\omega}^{\,\varsigma} \tilde{G}^{\phantom{\varsigma}}_{k'} \tilde{G}^{\phantom{\varsigma}}_{k'+q}  \Gamma_{\nu'\nu\omega}^{\,\varsigma}
\end{align}
We omit spin labels for Green's functions in this expression, because in the considered paramagnetic case the Green's function does not depend on the projection of spin. One can again use the partially bosonized representations for the four-point vertex function~\eqref{eq:Gamma}, which leads to
\begin{widetext}
\begin{align}
\tilde{\Sigma}_{k}^{(2)} = &-\frac14 \sum_{k', q}\Bigg\{ 4\sum_{\varsigma} \left(M^{\varsigma}_{\nu,\nu',\omega} M^{\varsigma}_{\nu',\nu,\omega} + M^{\varsigma}_{\nu,\nu+\omega,\nu'-\nu} M^{\varsigma}_{\nu',\nu'+\omega,\nu-\nu'} \right) + 4M^{\rm s}_{\nu,\nu',\nu+\nu'+\omega} M^{\rm s}_{\nu',\nu,\nu+\nu'+\omega} 
-4\,M^{\rm d}_{\nu,\nu',\omega}\,M^{\rm d}_{\nu',\nu'+\omega,\nu-\nu'} \notag\\
&+ 12\,M^{\rm m}_{\nu,\nu',\omega}\,M^{\rm m}_{\nu',\nu'+\omega,\nu-\nu'} 
- 24\,M^{\rm m}_{\nu,\nu',\omega}M^{\rm d}_{\nu',\nu'+\omega,\nu-\nu'} + 8\,M^{\rm d}_{\nu,\nu',\omega}M^{\rm s}_{\nu',\nu,\nu+\nu'+\omega} - 24\,M^{\rm m}_{\nu,\nu',\omega}M^{\rm s}_{\nu',\nu,\nu+\nu'+\omega} \Bigg\} \,\tilde{G}_{k'} \,\tilde{G}_{k'+q} \,\tilde{G}_{k+q}
\label{eq:Sigma2_app_full}
\end{align}
\end{widetext}
Shifting momentum and frequency indices as $k'\to{}k+q''$ and $q\to{}k''-k$ one can show that the product of two transverse fluctuations results in the same contribution to the self-energy as the product of two longitudinal ones 
\begin{align}
&\sum_{k',q} M^{\varsigma}_{\nu,\nu+\omega,\nu'-\nu} M^{\varsigma}_{\nu',\nu'+\omega,\nu-\nu'} \tilde{G}_{k'} \tilde{G}_{k'+q} \tilde{G}_{k+q} 
\to \notag\\
&\sum_{k'',q''} M^{\varsigma}_{\nu,\nu'',\omega''} M^{\varsigma}_{\nu+\omega'',\nu''+\omega'',-\omega''} \tilde{G}_{k+q''} \tilde{G}_{k''+q''} \tilde{G}_{k''} = \notag\\
&\sum_{k'',q''} M^{\varsigma}_{\nu,\nu'',\omega''} M^{\varsigma}_{\nu'',\nu,\omega''} \tilde{G}_{k''} \tilde{G}_{k''+q''} \tilde{G}_{k+q''}
\end{align}
The last relation can be obtained imposing the symmetry of the four- and three-point vertex functions~\eqref{eq:Gamma_appendix} and~\eqref{eq:Lambda_app}.
The product of two singlet fluctuations can also be simplified shifting $q\to{}q''-k-k'$ as
\begin{align}
&\sum_{k',q} M^{\rm s}_{\nu,\nu',\nu+\nu'+\omega} M^{\rm s}_{\nu',\nu,\nu+\nu'+\omega} \tilde{G}_{k'} \tilde{G}_{k'+q} \tilde{G}_{k+q} \to \notag\\ 
&\sum_{k',q''} M^{\rm s}_{\nu,\nu',\omega''} M^{\rm s}_{\nu',\nu,\omega''} \tilde{G}_{k'} \tilde{G}_{q-k} \tilde{G}_{q-k'}
\end{align}
Then, keeping only longitudinal contributions in $\tilde{\Sigma}_{k,\sigma_1}^{(2)}$, one gets the second-order self-energy of the $\text{D-TRILEX}$ approach
\begin{align}
\tilde{\Sigma}_{k,\sigma}^{(2)} \simeq 
&- \sum_{q, \varsigma} \tilde{G}_{k+q,\sigma} \Lambda^{\hspace{-0.05cm}\varsigma}_{\nu\omega}  \bar{w}^{\varsigma}_{\omega}  \left(\sum_{k',\sigma'}\Lambda^{\hspace{-0.05cm}*\,\varsigma}_{\nu'\omega} \tilde{G}_{k'\sigma'} \tilde{G}_{k'+q,\sigma'} \Lambda^{\hspace{-0.05cm}\varsigma}_{\nu'\omega}\right) \bar{w}^{\varsigma}_{\omega}  \Lambda^{\hspace{-0.05cm}*\,\varsigma}_{\nu\omega} \notag\\
&+
\sum_{q}\tilde{G}_{q-k,\overline{\sigma}} \Lambda^{\hspace{-0.05cm}\rm s}_{\nu\omega} \bar{w}^{\rm s}_{\omega} \left(-\sum_{k'}\Lambda^{\hspace{-0.05cm}*\,\rm s}_{\nu'\omega} \tilde{G}_{k'\uparrow} \tilde{G}_{q-k'\downarrow} \Lambda^{\hspace{-0.05cm}\rm s}_{\nu'\omega}\right) \bar{w}^{\rm s}_{\omega} \Lambda^{\hspace{-0.05cm}*\,\rm s}_{\nu\omega} \notag\\
=&-\sum_{q,\varsigma} \Lambda^{\hspace{-0.05cm}\varsigma}_{\nu\omega} \tilde{G}_{q+k,\sigma} \bar{w}^{\varsigma}_{\omega} \, \bar{\Pi}^{\varsigma}_{q} \, \bar{w}^{\varsigma}_{\omega} \Lambda^{\hspace{-0.05cm}*\,\varsigma}_{\nu\omega} \notag\\
&+ \sum_{q} \Lambda^{\hspace{-0.05cm}\rm s}_{\nu\omega} \tilde{G}_{q-k,\overline{\sigma}} \, \bar{w}^{\rm s}_{\omega} \bar{\Pi}^{\rm s}_{q} \bar{w}^{\rm s}_{\omega} \Lambda^{\hspace{-0.05cm}*\,\rm s}_{\nu\omega}
\label{eq:Sigma_2nd}
\end{align}
Neglecting non-longitudinal contributions in Eq.~\eqref{eq:Sigma2_app_full} explains the mismatch in the real part of the $\text{D-TRILEX}$ self-energy in the trivial regime of high-temperatures (${\beta=2}$) and weak interactions (${U=2}$). 

\subsubsection{Remaining part of the LDB self-energy}

The $\text{D-TRILEX}$ form of the remaining part of the LDB self-energy can be obtained preserving only longitudinal fluctuations in the partially bosonized representation for the four-point vertex~\eqref{eq:Gamma} ${\Gamma^{\,\varsigma}_{\nu\nu'\omega} \simeq 2 M^{\varsigma}_{\nu\nu'\omega}}$. Then, the renormalized three-point~\eqref{eq:Lvertex} and four-point vertices~\eqref{eq:Pvertex} become 
\begin{align}
L^{\hspace{-0.05cm}\varsigma}_{\nu q} &\simeq \Lambda^{\hspace{-0.05cm}\varsigma}_{\nu \omega}\left(1 + \bar{W}^{\varsigma}_{q}\bar{\Pi}^{\varsigma}_{q} \right)\\
{\rm P}^{\,\varsigma}_{\nu\nu'q} &\simeq 
2 \Lambda^{\hspace{-0.05cm}\varsigma}_{\nu\omega} \bar{W}^{\varsigma}_{q} \Lambda^{\hspace{-0.05cm}*\,\varsigma}_{\nu\omega}
\end{align}
where 
\begin{align}
\bar{W}_{q}^{\varsigma~-1} &= \bar{w}_{\omega}^{\varsigma~-1} - \bar{\Pi}^{\varsigma}_{q}
\end{align}
Substituting these expressions to the remaining part of the ladder contribution to the self-energy~\eqref{eq:Sigma_ladd}, one gets
\begin{align}
\tilde{\Sigma}^{(3+)}_{k\sigma} \simeq &-\sum_{q,\varsigma} \Lambda^{\hspace{-0.05cm}\varsigma}_{\nu\omega} \tilde{G}_{q+k,\sigma} \bar{W}^{\varsigma}_{q} \, \bar{\Pi}^{\varsigma}_{q} \, \bar{w}^{\varsigma}_{\omega} \, \bar{\Pi}^{\varsigma}_{q} \, \bar{w}^{\varsigma}_{\omega} \Lambda^{\hspace{-0.05cm}*\,\varsigma}_{\nu\omega} \notag\\
&+\sum_{q} \Lambda^{\hspace{-0.05cm}\rm s}_{\nu\omega} \tilde{G}_{q-k,\overline{\sigma}} \bar{W}^{\rm s}_{q} \, \bar{\Pi}^{\rm s}_{q} \, \bar{w}^{\rm s}_{\omega} \, \bar{\Pi}^{\rm s}_{q} \, \bar{w}^{\rm s}_{\omega} \Lambda^{\hspace{-0.05cm}*\,\rm s}_{\nu\omega}
\label{eq:Sigma_3rd}
\end{align}
In this expression we additionally introduced the screening of the four-point vertex in the particle-particle channel, which is usually not accounted for the LDF theory.
Combining all ladder terms~\eqref{eq:Sigma_1st},~\eqref{eq:Sigma_2nd}, and~\eqref{eq:Sigma_3rd} together, the LDF self-energy simplifies to
\begin{align}
\tilde{\Sigma}^{\rm LDF}_{k\sigma} 
&= -\sum_{q,\varsigma} \left\{ \Lambda^{\hspace{-0.05cm}\varsigma}_{\nu\omega} \tilde{G}_{q+k,\sigma} \bar{W}^{\varsigma}_{q} \Lambda^{\hspace{-0.05cm}*\,\varsigma}_{\nu\omega} 
- \Lambda^{\hspace{-0.05cm}\rm s}_{\nu\omega} \tilde{G}_{q-k,\overline{\sigma}} \bar{W}^{\rm s}_{q} \Lambda^{\hspace{-0.05cm}*\,\rm s}_{\nu\omega} \right\}
\end{align}
Under the same approximation, the mixed diagram~\eqref{eq:Sigma_mix} becomes
\begin{align}
\tilde{\Sigma}^{\rm mix}_{k\sigma} = &-\sum_{q,\varsigma} \Lambda^{\hspace{-0.05cm}\varsigma}_{\nu \omega} \tilde{G}_{q+k,\sigma} (1 + \bar{W}^{\varsigma}_{q}\bar{\Pi}^{\varsigma}_{q}) \tilde{W}^{\varsigma}_{q}(1+\bar{\Pi}^{\varsigma}_{q}\bar{W}^{\varsigma}_{q}) \Lambda^{\hspace{-0.05cm}*\,\varsigma}_{\nu \omega}\notag\\
&+\sum_{q} \Lambda^{\hspace{-0.05cm}\rm s}_{\nu \omega} \tilde{G}_{q-k,\overline{\sigma}} (1 + \bar{W}^{\rm s}_{q}\bar{\Pi}^{\rm s}_{q}) \tilde{W}^{\rm s}_{q}(1+\bar{\Pi}^{\rm s}_{q}\bar{W}^{\rm s}_{q}) \Lambda^{\hspace{-0.05cm}*\,\rm s}_{\nu \omega}
\end{align}
where we also added the contribution from the particle-particle channel. Using the Dyson equation~\eqref{eq:DualW_Dyson} for the dual bosonic propagator $\tilde{W}^{\vartheta}_{q}$ with the approximate dual polarization operator~\eqref{eq:DualPi} 
\begin{align}
\tilde{\Pi}^{\vartheta}_{q}
= \bar{\Pi}^{\vartheta}_{q} (1 + \bar{W}^{\vartheta}_{q}\bar{\Pi}^{\vartheta}_{q})
\end{align}
the total self-energy reduces to the $\text{D-TRILEX}$ result~\eqref{eq:Sigma_dual}
\begin{align}
\tilde{\Sigma}^{\rm D-TRILEX}_{k\sigma} &= \tilde{\Sigma}^{\rm LDF}_{k\sigma} + \tilde{\Sigma}^{\rm mix}_{k\sigma} \notag\\
&\simeq -\sum_{q,\varsigma} \left\{ \Lambda^{\hspace{-0.05cm}\varsigma}_{\nu\omega} \tilde{G}_{q+k,\sigma} W^{\varsigma}_{q} \Lambda^{\hspace{-0.05cm}*\,\varsigma}_{\nu\omega} 
- \Lambda^{\hspace{-0.05cm}\rm s}_{\nu\omega} \tilde{G}_{q-k,\overline{\sigma}} W^{\rm s}_{q} \Lambda^{\hspace{-0.05cm}*\,\rm s}_{\nu\omega} \right\}
\end{align}
The renormalized interaction of the theory can be found as follows
\begin{align}
W_{q}^{\vartheta~-1} &= {\cal W}_{q}^{\vartheta~-1} - \bar{\Pi}^{\vartheta}_{q} 
\end{align}
where the partially dressed bosonic propagator (see Eqs.~\eqref{eq:Wdm} and~\eqref{eq:Ws}) is 
\begin{align}
{\cal W}_{q}^{\vartheta} = \tilde{\cal W}_{q}^{\vartheta} + \bar{w}_{\omega}^{\vartheta}
\end{align}

\bibliography{main}

\end{document}